\documentclass[%
superscriptaddress,
amsmath,amssymb,
aps,pra,
twocolumn,
floatfix,
]{revtex4}

\usepackage{graphicx}
\usepackage{dcolumn}
\usepackage{bm}
\usepackage{hyperref}

\usepackage{amsmath,amsfonts,amssymb,amsthm}
\usepackage{braket}
\usepackage{slashed}
\usepackage{comment}
\usepackage{float}
\usepackage{mhchem}
\usepackage{tabularx}
\def \be {\begin{equation}}
\def \ee {\end{equation}}
\usepackage{times}
\usepackage{nicefrac}
\usepackage{epsf}
\usepackage{bm}
\usepackage{bbm}
\usepackage{color}
\usepackage{multirow}
\usepackage{booktabs}

\usepackage[labelformat=simple]{subcaption}

\DeclareCaptionLabelFormat{subcaptionlabel}{\normalfont(\textbf{#2}\normalfont)}
\preprint{APS/123-QED}
\begin{document}
\title{Insight into the prospective evaluation of third-order interelectronic corrections on Li-like ions}
\author{R. N. Soguel}
\email{romain.soguel@uni-jena.de}
\affiliation{Theoretisch-Physikalisches Institut, Friedrich-Schiller-Universität Jena, Max-Wien-Platz 1, 07743 Jena, Germany}
\affiliation{Helmholtz-Institut Jena, Fr\"obelstieg 3, 07743 Jena, Germany}
\affiliation{GSI Helmholtzzentrum für Schwerionenforschung GmbH, Planckstraße 1, 64291 Darmstadt, Germany}
\author{S. Fritzsche}
\affiliation{Theoretisch-Physikalisches Institut, Friedrich-Schiller-Universität Jena, Max-Wien-Platz 1, 07743 Jena, Germany}
\affiliation{Helmholtz-Institut Jena, Fr\"obelstieg 3, 07743 Jena, Germany}
\affiliation{GSI Helmholtzzentrum für Schwerionenforschung GmbH, Planckstraße 1, 64291 Darmstadt, Germany}
\date{\today}

\begin{abstract}
Relying on the redefined vacuum state approach, and based on one-particle three-loop Feynman diagrams, partial third-order interelectronic corrections to the valence electron energy shift are investigated in Li-like ions. The idea is to begin with simple one-particle gauge-invariant subsets composed of Feynman diagrams and to keep track of them in the many-electron frame, which is a strong asset of the formalism. An independent derivation is undertaken with the help of perturbation theory to cross-check the expressions. 
This two-method scheme helps to resolve how the different terms are distributed among three- and four-electron contributions. Furthermore, it provides a tool to overcome the difficulties related to the derivation of reducible terms, which are tricky to deal with. These two independent derivations and the comparison of the resulting expressions are fully consistent, except for two expressions. In these cases, the discrepancy can be traced back to a different topology of the poles.
\end{abstract}
\maketitle 

%
\section{Introduction}
Quantum electrodynamics (QED) is the prototype gauge theory on which the Standard Model (SM) of particles relies on. QED has shown to be impressively reliable in its ability to provide accurate predictions. To illustrate, consider the most precise prediction of the SM, the free electron magnetic moment in Bohr magnetons, $g/2$. A recent study measured its value to a spectacular precision of $1.3$ parts in $10^{13}$ \cite{Fan.2023_e-gfactor}. The SM prediction involves three sectors in its evaluation; it receives contributions from QED, as well as from the hadronic and weak interactions. For the former sector, the asymptotic power series in the fine-structure constant $\alpha$ is expanded up to the fifth order \cite{free_g-factor}, and contains muon and tauon contributions. 
The bound electron magnetic moment is more subtle to assess, nevertheless the stunning accuracy of $5.6$ parts in $10^{13}$ \cite{sailer2022measurement} was reached. To arrive at this value, a co-trapping of two different isotopes of neon ions was devised. Moreover, pushing QED in the presence of the binding nuclear field to its limits is a great way to gain in-depth knowledge about the theory and to probe potential new physics \cite{kozlov:2018}. 

Heavy highly charged ions offer a great natural laboratory to study QED in strong field regimes \cite{indelicato:2019:232001}. Unfortunately, such experimental high precision level is not achieved yet in the transition energies of heavy highly charged ions. Consider as an example the $1s$ Lamb shift in H-like ions, whose experimental value is 460.2 $\pm$ 4.6 eV  \cite{gumberidze:2005:223001}. The comparison between theory and experiment allows to probe first-order QED effects on a 1.7 percent level  \cite{indelicato:2019:232001}. In order to test second-order QED effects, which contribute -1.26(33)eV \cite{yerokhin:2003:073001}, demanding experimental updates are required \cite{gassner:2018:073033}. A part of the problem lies in the high energy of the transitions involved, where the sensitivity of the detectors is poor in the KeV regime. Note that a lot of energy is required to excite the tightest bound electron over the ionization threshold. A way to circumvent this issue is to probe many-electron transition energies, lying from soft x-ray to ultraviolet and, accessible by laser spectroscopy. Already for He-like ions, the uncertainties drop below the eV level \cite{amaro:2012:043005,Epp:2015:020502}. In the case of Li-like ions, the accuracy reached the meV level both in uranium  \cite{brandau:2003:073202,beiersdorfer:2005:233003} and xenon \cite{Bernhardt:2015:012710} ions. A similar accuracy was achieved in Be-like ions \cite{Bernhardt_2015}. Also in B-like \cite{draganic:2003:183001, mackel:2011:143002}, F-like \cite{lu:2020:042817, oneil:2020:032803} and Na-like \cite{Chen:2003:022507} ions precise measurements were conducted. To date, the most compelling experimental cases are found to be (i) the $2p_{3/2} - 2s$ transition in Li-like bismuth ions \cite{beiersdorfer:1998:3022}, (ii) the $2p_{1/2} - 2s$ transition in Li-like uranium ions \cite{beiersdorfer:2005:233003}. 

The increasing experimental precision drives theoretical predictions to their limits and enforces an accurate description of complex electrons dynamics. The evaluation of the dynamical properties and the structure of highly relativistic, tightly bound electrons in highly-charged ions with utmost accuracy represents one of the most important and demanding problems in modern theoretical atomic physics. In this view, the treatment of the interelectronic interaction is a fundament in order to achieve accurate theoretical predictions of the energy levels in many-electron atoms or ions. As a consequence, \textit{ab initio} calculations are the holy grail in the quest for many-electron atoms in the frame of bound-state QED (BSQED). The derivations performed so far used a zeroth-order many-electron wave function constructed as a Slater determinant (or sum of Slater determinants) with all electrons involved \cite{blundell:1993:2615, sapirstein:2015:062508, shabaev:1994:4489}. Such a derivation becomes increasingly difficult to handle for many-electron systems, especially when facing higher-order corrections. The framework of a vacuum state redefinition \cite{shabaev:2002:119, Soguel_2021pra,Soguel_2022} is proposed to tackle third-order interelectronic interactions. Such a technique is not yet widely spread in the BSQED community but already proved helpful in the evaluation of the screened radiative and two-photon exchange corrections to the $g$-factor and hyperfine splitting \cite{volotka:2009:033005, glazov:2010:062112, volotka:2012:073001, volotka:2014:253004, glazov:2019:173001, kosheleva:2020:013364}, the ground-state and ionization energies of Be-like ions \cite{malyshev:2014:062517, malyshev:2015:012514}, respectively, and for the transition energies between low-lying levels \cite{malyshev:2021:183001}. It showed to be of intrinsic relevance in the evaluation of the Delbrück scattering above the pair production threshold \cite{sommerfeldt2023allorder}. 

This work treats (a partial) third-order interelectronic correction to a single-valence state over closed-shells. The aim is to demonstrate that it is feasible to assess third-order interelectronic corrections even though such calculations are challenging and especially difficult. To exclude mistakes, and as a cross-check of the derived expressions, two different methods are utilised to obtain the results. The first one is an effective one-particle approach, which relies on the redefinition of the vacuum state. It deals with one-particle three-loop diagrams. Its presentation is given in Sec.~\ref{sec:third_order_rvs}, after a brief introduction to BSQED in Sec.~\ref{section:BSQED}. The idea is to provide a proof-of-principle that third-order interelectronic corrections can be tackled in the redefined vacuum state framework, owing to the transcription of the one-particle gauge-invariant (GI) subsets to many-electron diagrams. The second method, introduced in Sec.~\ref{sec:pert}, considers a perturbation theory approach to a two-photon-exchange subset. The subset involves a loop contribution, and a potential-like interaction accounts for the perturbation. Then, the results are mapped to the three-photon-exchange corrections. This independent derivation of the formulas ensures that potential mistakes are identified and ruled out. 
The resulting expressions contain infrared (IR) divergences. These are inspected in Sec.~\ref{sec:IR_div}, regularized by the introduction of a photon mass term, and it is shown that they are cancelled by terms within the proposed GI subsets. Section~\ref{sec:comparison} compares the outcomes of the two methods. 
The present work ends with a concluding discussion in Sec.~\ref{sec:discussion}.

Natural units ($\hbar = c = m_e = 1$) are used throughout this paper, the fine structure constant is defined as $\alpha = e^{2}/(4\pi),\,e < 0$. 
Unless explicitly stated, all integrals are implicitly assumed to be over the full real axis.


\section{Bound state QED}
\label{section:BSQED}
The relativistic quantum description of the electron-positron field is based on the Dirac equation. The framework of BSQED is based on the resummation of all Feynman diagrams involving the interaction of the (free) electron with the classical field of
the nucleus. Such a procedure, applied by Weinberg \cite{weinberg_1995}, leads to the bound-state electron propagator. Equivalently, from Furry's perspective \cite{furry:1951:115}, the interaction of the electron-positron field with the external classical field of the nucleus can also be taken into account non-perturbatively from the beginning by solving Dirac equation in the presence of the binding potential, 
\begin{equation} 
h_{\text{D}} \phi_j({\bf x}) = \left[-i \boldsymbol{\alpha} \cdot \boldsymbol{\nabla} + \beta  + V({\bf x})\right]\phi_j({\bf x})= \epsilon_j \phi_j({\bf x})\,,
\label{Dirac_eq}
\end{equation}
leading to the so-called Furry picture of QED. $\phi_j(\textbf{x})$ are the solutions of the stationary Dirac equation in the potential well $V({\bf x})$ occurring due to the nucleus and $j$ stands for all quantum numbers. $\alpha^{k}$ and $\beta$ are Dirac matrices , $V({\bf x}) = V_{\rm C}({\bf x})$ is the external classical Coulomb field arising from the nucleus. Solving Eq.~\ref{Dirac_eq} implies an all-order treatment in $\alpha Z$, with $Z$ the nuclear charge, hence going beyond perturbative regime. The time-depend solution is the stationary solution $\phi_k(\textbf{x})$ multiplied by the phase factor $\exp{\left(-i \epsilon_k t \right)}$. For completeness, the extended Furry picture accounts for the presence of a screening potential $U({\bf x})$ besides the Coulomb one, $V({\bf x}) = V_{\rm C}({\bf x}) + U({\bf x})$, which implies a partial consideration of interelectronic interactions. The redefinition of the vacuum state is conducted in such a way that all core orbitals from the closed shells belong to it \cite{lindgren_morrison}. The notation $\ket{\alpha}$ stands for
\begin{equation}
\ket{\alpha} = a_{a}^{\dagger} a_{b}^{\dagger} a_{c}^{\dagger} ... \ket{0}\,,
\label{redefined vacuum}
\end{equation}
and $\ket{\alpha}$ is going to be referred to as the redefined vacuum state. The following notation is applied, according to  Lindgren and Morisson \cite{lindgren_morrison} and Johnson \cite{johnson_2007}: $v$ designates a valence electron,  $a,b,c,...$ stands for core orbitals, $i,j,k,l,p,...$ correspond to arbitrary states. Upon second quantization, the (non-interacting) electron-positron field
can be expanded in terms of creation and annihilation operators. Within the redefined vacuum state approach, such decomposition still holds but needs to be slightly adapted with respect to the Fermi level $E_{\alpha}^{F}$,
\begin{equation}
\psi^{(0)}_{\alpha}(t,{\bf x}) = \sum_{\epsilon_j > E_{\alpha}^{F} } a_j \phi_j({\bf x}) e^{-i\epsilon_j t} + \sum_{\epsilon_j < E_{\alpha}^{F} } b_j^{\dagger } \phi_j({\bf x}) e^{-i\epsilon_j t}\,.
\label{psi_exp}
\end{equation}
The Fermi level of the redefined vacuum state lies between the highest core state and the valence state, $E_{\alpha}^{F}\in (\epsilon_a, \epsilon_v)$. Consequently, the electron propagator is affected in the following manner,
\begin{align}
\bra{\alpha}  T\left[\psi_{\alpha}^{(0)}(t,{\bf x})  \bar{\psi}_{\alpha}^{(0)}(t^\prime,{\bf y})\right] &\ket{\alpha} = \nonumber\\
= \frac{i}{2\pi} \int d\omega \sum_{j} &\frac{\phi_j({\bf x}) \bar{\phi}_j({\bf y}) e^{-i(t - t^\prime) \omega}}{\omega - \epsilon_j + i\varepsilon (\epsilon_j-E_{\alpha}^{F}) }\,,
\label{eq:e_propagator}
\end{align}
where $\epsilon>0$ implies the limit to zero. It is convenient to define $u=1-i \varepsilon$ for later use. 
The difference between the electron propagator in the redefined vacuum and in the standard one corresponds graphically to a cut of an inner electron line in the Feynman diagram. Such difference is mathematically implemented via the Sokhotski-Plemelj theorem. For $p \in \mathbb{N}^*$,
\begin{align}
\sum_j&\frac{\phi_j(\boldsymbol{x}) \bar{\phi}_j(\boldsymbol{y})}{\left[ \omega - \epsilon_j + i\varepsilon (\epsilon_j-E_{\alpha}^{F})\right]^p}
\nonumber \\ 
&-\sum_j\frac{\phi_j(\boldsymbol{x}) \bar{\phi}_j(\boldsymbol{y})}{\left[\omega - \epsilon_j + i\varepsilon (\epsilon_j-E^F)\right]^p}
\nonumber \\ 
=& \frac{2\pi i (-1)^p}{(p-1)!}\frac{d^{(p-1)}}{d \omega^{(p-1)}}\sum_{a} \delta(\omega - \epsilon_{a})\phi_{a}(\boldsymbol{x}) \bar{\phi}_{a}(\boldsymbol{y})\,,
\label{Sokhotski}
\end{align}
this equality is to be understood while integrating in the complex $\omega$-plane.
The reader is referred to Refs.~\cite{Soguel_2021pra, Soguel_2021sym, Soguel_2022} for more details on the vacuum state redefinition within the BSQED framework and the corresponding formalism.  

The light-unperturbed normal ordered Hamiltonian can be expressed as \cite{mohr:1998:227}
\begin{equation}
H_0 = \int d^{3}{\bf x} :\psi^{(0)\dagger}(t,{\bf x}) h_{\text{D}} \psi^{(0)}(t,{\bf x}):\,.
\end{equation}
It is left to introduce the light related part. The interaction Hamiltonian is constructed as
\begin{equation}
H_{\text{int}} = \int d^{3}{\bf x} :\psi^{(0)\dagger}(t,{\bf x}) h_{\text{int}} \psi^{(0)}(t,{\bf x}):\,.
\end{equation}
It encapsulates the interaction with the quantized electromagnetic field $A_{\mu}$ and the counterterm associated to the screening potential $-U({\bf x})$, when one works within the extended Furry picture. The explicitly expression for $h_{\text{int}}$ is $h_{\text{int}} = e \alpha^{\mu} A_{\mu}(t,{\bf x}) - U({\bf x})$. The effect of the interaction Hamiltonian is accounted for within BSQED perturbation theory. Different approaches are possible for its formulation \cite{mohr:1998:227, shabaev:2002:119, lindgren:2004:161, andreev:2008:135}. The calculation presented in what follows relies on the two-time Green's function (TTGF) method \cite{shabaev:2002:119}.

The photon propagator is denoted by $D_{\mu \nu} ({\bf x} - {\bf y};\omega)$, with $\omega$ the photon's energy. The interelectronic-interaction operator $I({\bf x}- {\bf y};\omega)$ is defined as $I({\bf x}- {\bf y}; \omega) = e^2 \alpha^{\mu} \alpha^{\nu} D_{\mu \nu} ({\bf x}- {\bf y}; \omega)$, where $\alpha^{\mu}  = (1, \boldsymbol{\alpha})$. The interelectronic-interaction matrix element $I_{ijkl}(\omega)$ is a shorthand notation standing for
\begin{equation}
I_{i j k l}(\omega) = \int d^3{\bf x} d^3{\bf y} \psi_i^{\dagger}({\bf x}) \psi_j^{\dagger}({\bf y}) I({\bf x} - {\bf y};\omega)  \psi_k({\bf x}) \psi_l({\bf y})\,,
\end{equation}
and satisfies the transposition symmetry property $I_{i j k l}(\omega) = I_{j i l k}(\omega)$. In the Feynman and Coulomb gauges, the interelectronic-interaction operator $I({\bf x}- {\bf y};\omega)$ is an even function of $\omega$.


\section{Redefined vacuum state approach}
\label{sec:third_order_rvs}
The idea behind a redefinition of the vacuum state is to benefit from a hydrogen-like picture of the problem, the system for which QED is the most developed. This setting is feasible as long as the system under consideration has a single valence electron above some closed shells. The essential notion in introducing a redefined vacuum state is to separate the electron dynamics into the “core” and “valence” parts. The first part is relegated to the reference vacuum energy and can be neglected when the transition energy -- with a significant many-electron background remaining unchanged -- is considered. The key feature of this approach is that the core contributions -- arising from the interaction between core electrons are canceled in the difference between the excited and the ground state energies -- are not considered from the very
beginning.

The effective one-particle approach, based on a redefinition of the vacuum state, is applied here to more involved Feynman diagrams, as a proof-of-principle that advanced calculation can be undertaken in this way. The three-photon-exchange corrections investigated, as well as the specific one-particle three-loop Feynman diagrams serving as a starting point of the subsequent consideration were selected with hindsight. It was necessary to be able to verify, in some manner, the obtained expressions. The perturbative approach requires already known diagrams, therefore, some totally new topology in the considered diagrams was not a realistic choice such as for example, a one-loop correction shared among three electrons exists. Hence, after the introduction of the necessary elements of the TTGF method, a partial recap on the two-photon-exchange correction is undertaken. It both details how the investigated diagrams were selected and serves as a staring point for later developments. 

The investigation is carried out for the valence state described as $\ket{v} = a_v^{\dagger} \ket{\alpha}$ in the perspective of a redefined vacuum state. The cornerstone expression for the energy shift is  
\begin{equation}
\Delta E_v = \frac{\displaystyle{\frac{1}{2\pi i}} \oint_{\Gamma_v} dE (E-\bra{v} H_0 \ket{v}) \bra{v}\Delta g_{\alpha}(E)\ket{v}} {1 + \displaystyle{\frac{1}{2\pi i}} \oint_{\Gamma_v} dE \bra{v}\Delta g_{\alpha}(E)\ket{v}}\,,
\label{energy correction}
\end{equation}
where the contour $\Gamma_v$ is chosen such that it surrounds anticlockwise only the pole $E^{(0)} = \bra{v} H_0 \ket{v} \equiv \epsilon_v$. Other singularities are kept outside this contour. The Fourier transformed TTGF matrix element is provided by $\bra{v} \Delta g_{\alpha} (E)\ket{v} = \Delta g_{\alpha, vv}$, where $\Delta g_{\alpha}(E)$ stands for $\Delta g_{\alpha}(E)= g_{\alpha}(E) - g_{\alpha}^{(0)}(E)$ and $g_{\alpha}^{(0)}(E)$ is the zeroth-order Fourier transformed TTGF. The Fourier transformed TTGF, where the coordinates have been integrated out, is given by
\begin{align}
g_{\alpha}&(E) \delta(E-E^{\prime}) = \frac{1}{2\pi i} \sum_{i,j}\int d^{3}x d^{3}y \int dt dt^{\prime} e^{i (E t -E^{\prime}t^{ \prime})} \nonumber\\
&\times \phi^{\dagger}_i(\boldsymbol{x})\bra{\alpha} T\left[ \psi_{\alpha} (t, \boldsymbol{x}) \psi_{\alpha}^{\dagger} (t^{\prime}, \boldsymbol{y}) \right] \ket{\alpha}\phi_j(\boldsymbol{y})a_i^{\dagger}a_j\,.
\end{align}
 The treatment of the Green's function within perturbation theory allows to expend it to the different orders in $\alpha$: $\Delta g_{\alpha}(E) =\Delta g_{\alpha}^{(1)}(E) + \Delta g_{\alpha}^{(2)}(E) + \Delta g_{\alpha}^{(3)}(E) + \cdots$. Isolating the corresponding third order, the resulting expression for the energy shift can be represented as
\begin{widetext}
\begin{align} 
\Delta E^{(3)}_v =& \, \frac{1}{2\pi i}\oint_{\Gamma_v} dE (E-\epsilon_v) \Delta g_{\alpha,vv}^{(3)}(E) 
-  \frac{1}{2\pi i}\oint_{\Gamma_v} dE (E-\epsilon_v) \Delta g_{\alpha,vv}^{(2)}(E)  \frac{1}{2\pi i}\oint_{\Gamma_v} dE^{\prime} \Delta g_{\alpha,vv}^{(1)}(E^{\prime})  \nonumber \\
&-  \frac{1}{2\pi i}\oint_{\Gamma_v} dE (E-\epsilon_v) \Delta g_{\alpha,vv}^{(1)}(E)
 \left\{ \frac{1}{2\pi i}\oint_{\Gamma_v} dE^{\prime} \Delta g_{\alpha,vv}^{(2)}(E^{\prime}) -  \left[   \frac{1}{2\pi i}\oint_{\Gamma_v} dE^{\prime}  \Delta g_{\alpha,vv}^{(1)}(E^{\prime}) \right]^2  \right\} \,.
\label{3rd order}
\end{align}
\end{widetext}
The terms which do not involve $\Delta g_{\alpha,vv}^{(3)}(E)$ are referred to as disconnected elements. 
A peculiar attention is required in the treatment of the contributions where the energy of an intermediate state equals to the energy of the reference state. These type of contributions are so-called \textit{reducible} terms.  Three different type of Feynman diagrams are found at this order and are separated accordingly as follows: the one-particle loop diagrams are denoted by (L), the interelectronic-interaction diagrams by (I) and the screened-loop diagrams by (S). This reports focuses on the third-order interelectronic interactions (I), also referred to as three-photon-exchange corrections.

\subsection{Partial recap on two-photon-exchange corrections}

To showcase that the vacuum state redefinition approach is working, one resorts to results from Ref.~\cite{Soguel_2021pra}. The two-photon-exchange corrections were derived within the framework of a vacuum state redefinition, thus starting from an effective one-particle picture.

\begin{figure}[h]
\centering
\includegraphics[width=0.47\textwidth]{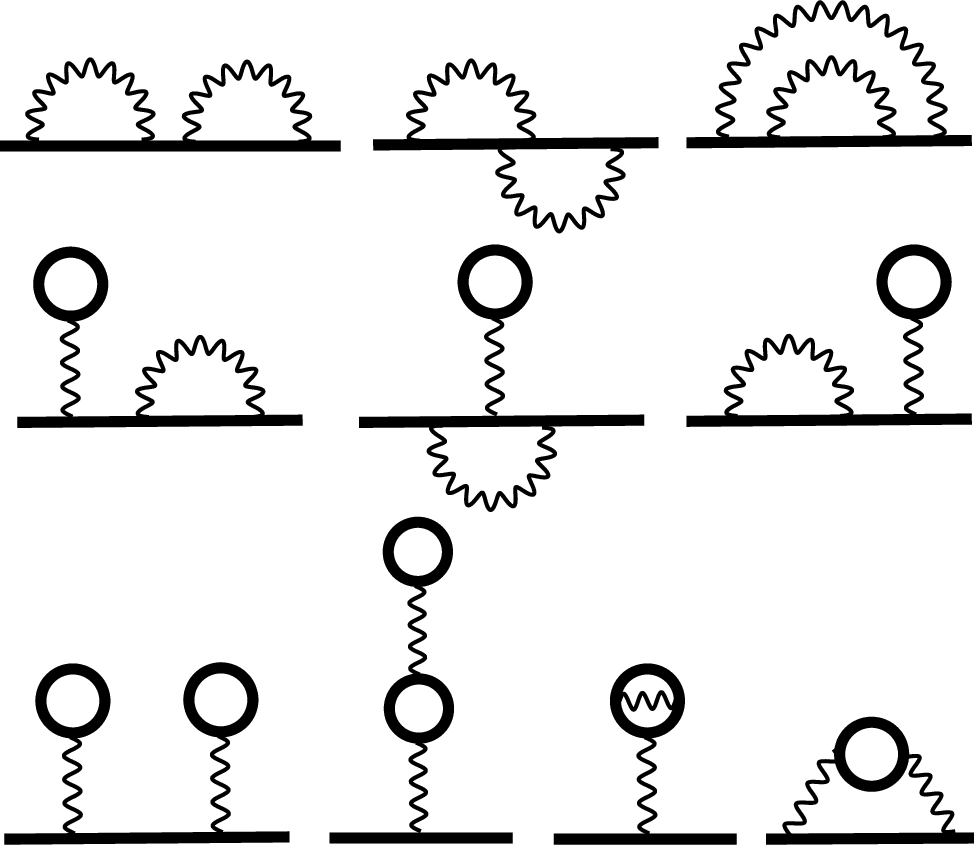}
\caption{The second-order one-electron Feynman diagrams labeled as follows: SESE (first row); SEVP (second row); VPVP, V(VP)P, V(SE)P, and S(VP)E from left to right in the last row. Thick black lines denote the electron propagator in the redefined vacuum state in an external potential $V$. Wavy lines represent to the photon propagator. }
\label{fig:1}
\end{figure}

\noindent The complete set of second-order one-particle diagrams consists of ten two-loop diagrams. These are presented in Fig. \ref{fig:1}. The gauge invariance of the one-particle two-loop diagrams was shown in Ref.~\cite{yerokhin:2006:253004}.  Eight gauge invariant subsets are identified based on the decomposition provided in this paper. The identified subsets should be gauge invariant in both the redefined and the standard vacuum state frameworks. This means that the many-electron diagrams obtained as a difference between redefined and standard vacuum state diagrams can be also classified according to these subsets. The subsets are labelled according to their composition in radiative-loop corrections. In what follows, SE stands for the \textit{self-energy} loop and VP for the \textit{vacuum-polarization} loop. The subsets, with the labeling presented in Figs. \ref{fig:1}, are the SESE one in the first line, the SEVP one in the second line, and the VPVP, V(VP)P, V(SE)P, and S(VP)E ones, from left to right, in the third line.  The eight identified GI subsets in the original Furry picture are, in terms of the one-electron description: SESE (two- and three-electron subsets), SEVP, S(VP)E (two- and three-electron subsets), VPVP, V(VP)P, and V(SE)P.

The resulting two-photon-exchange corrections are divided into three categories, in the many-electron frame; the ladder diagram (referred to as ladder loop), the crossed-ladder diagram (referred to as the crossed loop) and the three-electron diagram (referred to as the three-electron term in this subsection).

It was shown that that the ladder- and crossed-loop arose from the SESE and S(VP)E subsets \cite{comment_Soguel}. The SESE subset generates the exchange parts of the two-photon-exchange corrections, whereas the S(VP)E subset generates the direct parts of the two-photon-exchange corrections. The decision was made to pick out the S(VP)E subset as a starting point of the third-order interelectronic corrections. The reasons are (i) its simplicity in terms of the number of diagrams to consider (3 vs 1), and (ii) the fact that dealing with the exchange part is more difficult. The one-particle three-loop Feynman diagrams are given in Figs.~\ref{fig:H_subset} [S[V(VP)P]E subset] and \ref{fig:F_subsets} [S(VP)EVP subset]. They correspond to all possible insertion of a VP loop in the S(VP)E graph, being the simplest extension towards a one-loop three-photon-exchange corrections. Note that the inclusion of an extra energy-independent interelectronic operator, arising from the cut in the inserted VP loop, does not spoil gauge invariance. The one-photon-exchange operator was shown to be gauge invariant \cite{Soguel_2021pra}. An important point to highlight is that the $\omega$-flow in the VP loop must be symmetrized in the integrals of the final expressions in order to achieve gauge invariance. It was a crucial step when dealing with the S(VP)E subset (as shown below). The equality (\ref{eq:pole_symm}), and its third-order-pole version, were applied to the derived expressions in order to compare them with those based on the perturbation theory approach.

The expressions related to the S(VP)E subset are displayed below, as the initial expressions for later derivations based on perturbation theory. The proof for the gauge invariance of the S(VP)E subset is given numerically, for the Feynman and Coulomb gauges, in Tables I and II at the corresponding lines, and analytically in Eq.~(67) for the three-electron terms in Ref.~\cite{Soguel_2021pra}. The two-electron irreducible part $\Delta E^{(2\text{I})\text{S(VP)E,2e,irr}}_{v}$ reads
\begin{align}
\Delta& E^{(2\text{I})\text{S(VP)E,2e,irr}}_{v} = \nonumber \\
&\frac{i}{2\pi}\int d\omega \left[  \sum_{a, i, j}^\prime \frac{I_{v j i a}(\omega) I_{i a v j}( \omega)}{(\epsilon_v- \omega - \epsilon_{i} u) (\epsilon_a - \omega - \epsilon_{j} u)} \right. \nonumber \\
& \left. + \sum_{a, i, j}^{(i,j) \neq (a,v)} \frac{ I_{v a i j }(\omega ) I_{i j v a}( \omega)}{(\epsilon_v- \omega - \epsilon_{i} u) (\epsilon_a + \omega - \epsilon_{j} u)} \right]\,,
\label{eq:S(VP)E,2e,irr}
\end{align}
where one excludes the contribution with $\epsilon_i = \epsilon_v$ and $\epsilon_j = \epsilon_a$ from the crossed-direct term (first item) by hand and note this by the prime on the sum. The two-electron reducible part $\Delta E^{(2\text{I})\text{S(VP)E,2e,red}}_{v}$, coming from the ladder direct restriction $\epsilon_{i} + \epsilon_{j} = \epsilon_a + \epsilon_v$ together with the excluded part from the irreducible crossed-direct term, yields
\begin{align}
\Delta &E^{(2\text{I})\text{S(VP)E,2e,red}}_{v} = \nonumber \\ 
&-\frac{i}{4\pi} \int d\omega \left[ \frac{1}{(\omega + i \varepsilon)^2}+ \frac{1}{(-\omega + i \varepsilon)^2} \right] \nonumber\\ 
&\times \sum_{a, a_1, v_1} I_{v a a_1 v_1 }(\Delta_{va}-\omega) I_{a_1 v_1 v a}(\Delta_{va}-\omega) \,.  
\label{eq:S(VP)E,2e,red}
\end{align}
%
%
The three-electron irreducible part $\Delta E^{(2\text{I})\text{S(VP)E,3e,irr}}_{v}$ yields 
\begin{align} 
\Delta&E^{(2\text{I})\text{S(VP)E,3e,irr}}_{v} = \nonumber \\
&-\sum_{a,b,i}^{(i, b) \neq (v, a)}  \frac{ I_{v a b i}(\Delta_{vb})  I_{b i v a}(\Delta_{vb})  + I_{v a i b}(\Delta_{ba})  I_{i b v a}(\Delta_{ba})  }{(\epsilon_v +\epsilon_a -\epsilon_b -\epsilon_{i} ) } \nonumber \\ 
&-\sum_{a,b,i}  \frac{ I_{v i b a}(\Delta_{vb})  I_{b a v i}(\Delta_{vb})    }{(\epsilon_a +\epsilon_b -\epsilon_v -\epsilon_{i} )}\,,
\label{eq:S(VP)E,3e,irr}
\end{align}
together with the corresponding reducible part, 
\begin{equation}
\Delta E^{(2\text{I}) \text{S(VP)E,3e,red}}_{v} = -\sum_{a, a_1, v_1}
I_{v a a_1 v_1}(\Delta_{va}) I^{\prime}_{a_1 v_1 v a}(\Delta_{va})\,. 
\label{eq:S(VP)E,3e,red}
\end{equation}

As a side remark, note that the identity 
\begin{equation}
\frac{-1}{(x+ i\varepsilon)^2}  + \frac{1}{(-x+ i\varepsilon)^2} = \frac{2\pi}{i} \partial_x \delta(x)    
\label{eq:pole_symm}
\end{equation}
was used to symmetrize the $\omega$-flow in the VP loop, or in other words, to symmetrize the pole structure with respect to the real-line axis.


\begin{widetext}

\subsection{S[V(VP)P]E subset}
%
\begin{figure}[h]
\centering
\includegraphics[width=0.45\textwidth]{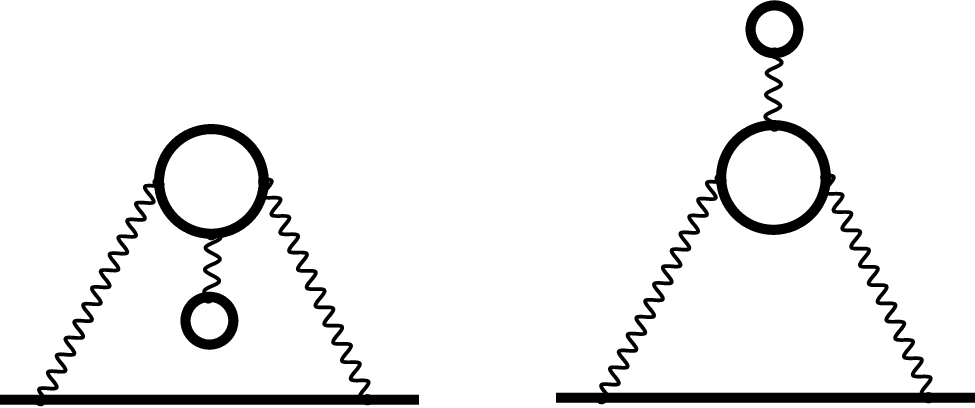}
\caption{One-particle three-loop Feynman diagrams of the S[V(VP)P]E subset participating to the third-order contribution to the energy shift of a single-particle state. They are denoted by $H_1$ (left) and $H_2$ (right). Notations are the same as in Fig. \ref{fig:1}.}
\label{fig:H_subset}
\end{figure} 
%
The Green's function corresponding to each diagram in the subset are given by
\begin{align} 
\Delta g^{(3) H_1}_{\alpha,\, vv}(E) =& \,  \frac{1}{(E - \epsilon_v)^2} \left( \frac{i}{2\pi} \right)^3 \sum_{i, j, k, l, p} \int d\omega dk_1 dk_2 \frac{ I_{v p i k}(\omega) }{[E - \omega - \epsilon_{i} + i\varepsilon (\epsilon_i - E_\alpha^F)]  [k_1 - \epsilon_{j} + i\varepsilon (\epsilon_j - E_\alpha^F)] } \nonumber \\
&\times \frac{ I_{k j l j}(0)  I_{l i p v}(\omega)  }{ [k_2  - \epsilon_{k} + i\varepsilon (\epsilon_k - E_\alpha^F)] [k_2 -  \epsilon_{l} + i\varepsilon (\epsilon_l - E_\alpha^F)]  [k_2 - \omega - \epsilon_{p} + i\varepsilon (\epsilon_p - E_\alpha^F)]} \,, 
\end{align}
for $H_1$ and by
\begin{align}
\Delta g^{(3) H_2}_{\alpha,\, vv}(E) =& \,  \frac{1}{(E - \epsilon_v)^2} \left( \frac{i}{2\pi} \right)^3 \sum_{i, j, k, l, p} \int d\omega dk_1 dk_2  \frac{ I_{v k i j}(\omega) }{ [E - \omega - \epsilon_{i} + i\varepsilon (\epsilon_i - E_\alpha^F)] [k_1 - \epsilon_{j} + i\varepsilon (\epsilon_j - E_\alpha^F)] } \nonumber \\
&\times \frac{ I_{l p k p}(0)  I_{j i l v}(\omega)    }{  [k_1 -\omega - \epsilon_{k} + i\varepsilon (\epsilon_k - E_\alpha^F)] [k_1 -\omega -  \epsilon_{l} + i\varepsilon (\epsilon_l - E_\alpha^F)]  [k_2  - \epsilon_{p} + i\varepsilon (\epsilon_p - E_\alpha^F)] } \,.
\end{align}
\end{widetext}
for $H_2$. According to the line of reasoning presented in the treatment of the SESE subset in Ref.~\cite{Soguel_2021pra}, starting from the previous Green's functions, the extraction of the different three-photon-exchange corrections is carried out with the subtraction of the corresponding expression in the standard vacuum state
\begin{equation}
\Delta E_v^{(3)}- \Delta E_v^{(3\text{L})} = \Delta E_v^{(3\text{I})}+ \Delta E_v^{(3\text{S})} \,.    
\label{eq:extraction}
\end{equation}
In other words, within the framework of a redefined vacuum state, the interelectronic and the screened corrections are treated on the same footing. As the diagrams are one-particle irreducible (1PI), one does not need to worry about the disconnected elements in Eq.~(\ref{3rd order}). Note that double cuts are possible in both diagrams, leading to so called non-diagrammatic elements. These are important to consider in order to properly treat the reducible part. Non-diagrammatic elements are expressions that cannot be drawn as single Feynman diagrams since they involve reducible parts. The results are separated in three- and four-electron contributions. 

To illustrate the extraction procedure based on Eq.~(\ref{eq:extraction}), the term-by-term three-electron (loop diagram) expressions obtained afterwards are displayed explicitly below for $H_1$. 
One distinguishes between three different type of terms: irreducible (irr), reducible 1 (red1) and reducible 2 (red2). The irreducible terms correspond to a first order pole ($S$-matrix terms) in the first expression of Eq.~(\ref{3rd order}), whereas the reducible 1 and 2 terms correspond to a second and third order pole, respectively, in the first expression of Eq.~(\ref{3rd order}). To keep track of the source of generated reducible parts, a subscript is used with previous notation; for example, $v_1$, $a_1$, $b_2$, where $\epsilon_{i_1} = \epsilon_i$. The different terms corresponding to crossed graphs are found to be
\begin{widetext}

%
\begin{equation} 
\Delta E^{(3\text{I})\text{3e,cross,irr}}_{v,\,H_1} = \frac{i}{2\pi} \int d\omega \sum_{i,j,k}^{k\neq b}  \frac{I_{v j i b}(\omega)  I_{b a k a}(0)  I_{ k i  j v}(\omega)  }{ (\epsilon_v -\omega -\epsilon_i u) (\epsilon_b -\omega -\epsilon_j u) (\epsilon_b -\epsilon_k u)  }\,,
\label{h1_cross_irr_7}
\end{equation}
\begin{equation} 
\Delta E^{(3\text{I})\text{3e,cross,irr}}_{v,\,H_1} = \frac{i}{2\pi} \int d\omega \sum_{i,j,k}^{k\neq b}  \frac{I_{v j i k}(\omega)  I_{k a b a}(0)  I_{b i j v}(\omega)  }{ (\epsilon_v -\omega -\epsilon_i u) (\epsilon_b -\omega -\epsilon_j u) (\epsilon_b -\epsilon_k u) }\,.
\label{h1_cross_irr_8}
\end{equation}
The terms associated to ladder graphs are given as follows,
\begin{equation} 
\Delta E^{(3\text{I})\text{3e,lad,irr}}_{v,\,H_1} = \frac{i}{2\pi} \int d\omega \sum_{i,j,k}^{ \{i,j \} \neq \{v,b \}, \{i,k \} \neq \{v,b \}}  \frac{I_{v b i j}(\omega)  I_{j a k a}(0)  I_{k i bv}(\omega)  }{ (\epsilon_v -\omega -\epsilon_i u) (\epsilon_b + \omega -\epsilon_j u) (\epsilon_b +\omega -\epsilon_k u) }\,,
\end{equation}
\begin{equation} 
\Delta E^{(3\text{I})\text{3e,lad,red1}}_{v,\,H_1} = -\frac{i}{2\pi} \int d\omega \sum_{i,j,k}^{ \{i,j \} = \{v,b \}, \{i,k \} \neq \{v,b \}}  \frac{I_{v b i j}(\omega)  I_{j a k a}(0)  I_{k i bv}(\omega)  }{ (\epsilon_v -\omega -\epsilon_i u)^2  (\epsilon_b +\omega -\epsilon_k u) }\,,
\label{h1_lad_red1_10}
\end{equation}
\begin{equation} 
\Delta E^{(3\text{I})\text{3e,lad,red1}}_{v,\,H_1} = -\frac{i}{2\pi} \int d\omega \sum_{i,j,k}^{ \{i,j \} \neq \{v,b \}, \{i,k \} = \{v,b \}}  \frac{I_{v b i j}(\omega)  I_{j a k a}(0)  I_{k i bv}(\omega)  }{ (\epsilon_v -\omega -\epsilon_i u)^2 (\epsilon_b + \omega -\epsilon_j u)  }\,,
\label{h1_lad_red1_11}
\end{equation}
\begin{equation} 
\Delta E^{(3\text{I})\text{3e,lad,red2}}_{v,\,H_1} = \frac{i}{2\pi} \int d\omega \sum_{i,j,k}^{ \{i,j \} = \{v,b \}, \{i,k \} = \{v,b \}}  \frac{I_{v b i j}(\omega)  I_{j a k a}(0)  I_{k i bv}(\omega)  }{ (\epsilon_v -\omega -\epsilon_i u)^3 }\,.
\label{h1_lad_red2}
\end{equation}
The non-diagrammatic element for $H_1$ reads
\begin{equation} 
\Delta E^{(3\text{I})\text{3e,cross,red1}}_{v,\,H_1} = -\frac{i}{2\pi} \int d\omega \sum_{i,j} \frac{I_{v j i b}(\omega)  I_{b a b_1 a}(0)  I_{b_1 i j v}(\omega)  }{ (\epsilon_v -\omega -\epsilon_i u)  (\epsilon_b -\omega -\epsilon_j u)^2 }\,.
\label{h1_cross_red1}
\end{equation}
%
The expressions presented above are the bare results after carrying out the difference of the vacuum states to extract the interelectronic interactions. We emphasis that the reducible expressions are IR divergent. These IR divergences are extracted and regularized later on, such as the symmetrization of the poles with regard to the real axis. The term-by-term interelectronic expressions related to $H_2$ are given in Appendix \ref{SVVPPE}. They are needed to show the explicit cancellation of IR divergences at the single Feynman diagram level. The final interelectronic three-electron expressions are found in Appendix \ref{sec:third-order_3e}, when the comparison with the results obtained from the perturbation theory approach is made. The four-electron terms are displayed in Appendix \ref{sec:third-order_4e}, again for a comparison with the outcome of the second method.

\subsection{S(VP)EVP subset}
\begin{figure}[h]
\centering
\includegraphics[width=0.45\textwidth]{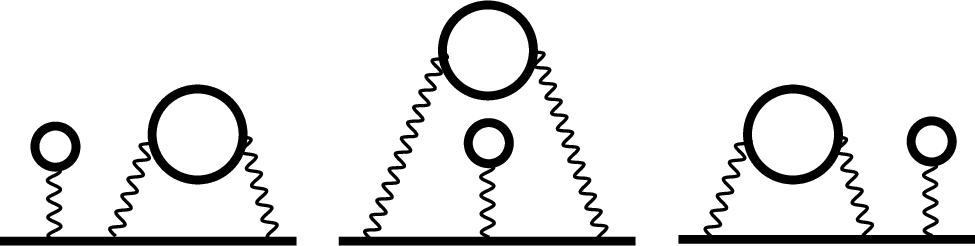}
\caption{One-particle three-loop Feynman diagrams of the S(VP)EVP subset participating to the third-order contribution to the energy shift of a single-particle state. They are denoted by $F_1$ (left), $F_2$ (middle) and $F_3$ (right). Notations are similar to Fig.~\ref{fig:1}.}
\label{fig:F_subsets}
\end{figure} 
As a second example, consider the Green's function associated to $F_2$ diagram
\begin{align}
\Delta g^{(3) F_2}_{\alpha,\, vv}(E) =&\, \frac{1}{(E - \epsilon_v)^2} \left( \frac{i}{2\pi} \right)^3 \sum_{i, j, k, l, p} \int d\omega dk_1 dk_2 \frac{ I_{v l i l}(\omega)}{[E - \omega - \epsilon_{i} + i\varepsilon (\epsilon_p - E_\alpha^F)] [k_1 - \epsilon_{j} + i\varepsilon (\epsilon_j - E_\alpha^F)]} \nonumber \\
&\times \frac{I_{i j k j}(0)   I_{l k p v}(\omega)   }{  [E - \omega - \epsilon_{k} + i\varepsilon (\epsilon_p - E_\alpha^F)]  [k_2 -  \epsilon_{l} + i\varepsilon (\epsilon_l - E_\alpha^F)] [k_2 -\omega - \epsilon_{p} + i\varepsilon (\epsilon_p - E_\alpha^F)]}\,. 
\end{align}
\end{widetext}
The identical procedure as stated above is applied to infer the three-photon-exchange corrections. The diagrams are not 1PI, with the exception of $F_2$. Therefore, one has to consider the second-order Green's function $\Delta g_\alpha^{(2) \text{S(VP)E}}$  and  the first-order one $\Delta g_\alpha^{(1) \text{VP}}$ to evaluate the disconnected elements in Eq.~(\ref{3rd order}). It turns out that only the first part of the fourth line survives. The disconnected elements are important to account for as they remove the redundancies in the reducible parts. To illustrate, $F_1$ and $F_3$ give redundant reducible terms, but the disconnected elements remove the extra ones, coming from $F_1$ in this case. A small subtlety is nevertheless present if an extra pole is explicitly extracted. In such a case, this reducible term is kept for two reasons. First  it cannot be generated in the disconnected ones, and second, which is more important, to achieve the IR finiteness of the expressions in the subset. Hence, for the S(VP)EVP subset, on the top of the peculiar extra-pole-term arising from $F_1$, only the $F_2$ and $F_3$ reducible IR divergent expressions are displayed in Table \ref{tab:IR_div}. Non-diagrammatic elements are only present in the $F_2$ diagram, and are of four-electron type. The results are separated in three- and four-electron contributions. The Feynman diagram $F_2$ leads to the subsequent expressions. Only an expression associated to a crossed-loop graph is found 
\begin{widetext}

\begin{equation}
\Delta E^{(3\text{I}) \text{3e,cross}}_{v,\,F_2} = \frac{i}{2\pi} \int d\omega \sum_{i,j,k}   \frac{ I_{v k i b}(\omega)  I_{i a j a}(0)  I_{b j k v}(\omega)    }{(\epsilon_v - \omega - \epsilon_i u) (\epsilon_v - \omega - \epsilon_j u) (\epsilon_b -\omega - \epsilon_k u)}\,,
\label{F2_cross}
\end{equation}
so it is for the ladder-loop graph, but incorporating the different types  of terms  (irr, red1 and red2) stated above  
\begin{equation} 
\Delta E^{(3\text{I}) \text{3e,lad,irr}}_{v,\,F_2} = \frac{i}{2\pi} \int d\omega \sum_{i,j,k}^{ \{i,k\} \neq \{v,b\}, \{j,k\} \neq \{v,b\} }   \frac{ I_{v b i k}(\omega)  I_{i a j a}(0)  I_{k j b v}(\omega)    }{(\epsilon_v - \omega - \epsilon_i u) (\epsilon_v - \omega - \epsilon_j u) (\epsilon_b +\omega - \epsilon_k u)}\,,
\end{equation}
\begin{align}
\Delta E^{(3\text{I}) \text{3e,lad,red1}}_{v,\,F_2} =& -\frac{i}{2\pi} \int d\omega \sum_{i,j,k}^{ \{i,k\} = \{v,b\}, \{j,k\} \neq \{v,b\} }  \left\{ \frac{ I_{v b i k}(\omega)  I_{i a j a}(0)  I_{k j b v}(\omega)    }{(\epsilon_v - \omega - \epsilon_j u)^2 } \left[  \frac{1}{(\epsilon_v - \omega - \epsilon_i u)} + \frac{1}{(\epsilon_b +\omega - \epsilon_k u)} \right]  \right. \nonumber  \\
&+ \left.\frac{ I_{v b i k}(\omega)  I_{i a j a}(0)  I_{k j b v}(\omega)    }{(\epsilon_v - \omega - \epsilon_i u)^2 (\epsilon_v - \omega - \epsilon_j u) }  \right\}\,,
\label{F2_lad_red11}
\end{align}
\begin{align} 
\Delta E^{(3\text{I}) \text{3e,lad,red1}}_{v,\,F_2} =& -\frac{i}{2\pi} \int d\omega \sum_{i,j,k}^{ \{i,k\} \neq \{v,b\}, \{j,k\} = \{v,b\} }  \left\{ \frac{ I_{v b i k}(\omega)  I_{i a j a}(0)  I_{k j b v}(\omega)    }{(\epsilon_v - \omega - \epsilon_i u)^2 } \left[  \frac{1}{(\epsilon_v - \omega - \epsilon_j u)}  +  \frac{1}{(\epsilon_b +\omega - \epsilon_k u)} \right] 
\right. \nonumber \\
&+ \left. \frac{ I_{v b i k}(\omega)  I_{i a j a}(0)  I_{k j b v}(\omega)    }{(\epsilon_v - \omega - \epsilon_i u) (\epsilon_v - \omega - \epsilon_j u)^2 } \right\} \,,
\label{F2_lad_red12}
\end{align}
\begin{align} 
\Delta E^{(3\text{I}) \text{3e,lad,red2}}_{v,\,F_2} =& \, \frac{i}{2\pi} \int d\omega \sum_{i,j,k}^{ \{i,k\} = \{v,b\}, \{j,k\} = \{v,b\} } \left\{ -\frac{I_{v b i k}(\omega)  I_{i a j a}(0)  I_{k j b v}(\omega)}{ (\epsilon_v - \omega - \epsilon_i u) (\epsilon_v - \omega - \epsilon_j u) } \left [  \frac{1}{(\epsilon_v - \omega - \epsilon_i u)} + \frac{1}{ (\epsilon_v - \omega - \epsilon_j u)} \right ]  \nonumber \right.\\
&+ \left. \frac{ I_{v b i k}(\omega)  I_{i a j a}(0)  I_{k j b v}(\omega)    }{(\epsilon_v - \omega - \epsilon_i u)^3 }  \right\}
\,.
\label{F2_lad_red2}
\end{align}
\end{widetext}
As in the case of the S[V(VP)P]E subset, the expressions showed above are those obtained in the extraction procedure of the interelectronic interaction, as the difference of the vacuum states. Some more work is required to extract the IR divergences of the reducible terms, to regularize them, as well as to symmetrize the poles with regard to the real axis. The expressions associated to $F_1$ and $F_3$ diagrams are found in Appendix \ref{SVPEVP}. The explicit cancellation of IR divergences at the single diagram level requires them. The final expressions associated to the one-loop three-photon-exchange diagrams are found in Appendix \ref{sec:third-order_3e}, when comparing with the results from the second derivation. The four-electron terms are displayed, for a comparison with the perturbation theory approach, in Appendix \ref{sec:third-order_4e}.


\section{Perturbation theory approach }
\label{sec:pert}
The idea behind this approach is to perturb a two-photon-exchange correction by the presence of some potential-like interaction  $\mathcal{V}$, to its first order. This method was applied in Ref.~\cite{Yerokhin_2021_104} to generate the two-photon-exchange corrections to the $g$-factor of Li-like ions. Specifically, the one-electron external wave function is perturbed as  
\begin{align}
\ket{i} \rightarrow \ket{i} + \ket{ \delta i}, \hspace{10pt}  \ket{ \delta i} = \sum_{j\neq i} \frac{\ket{j}  \mathcal{V}_{ji}  }{\epsilon_i - \epsilon_j} \,,
\label{eq:pert_state}
\end{align}
the energy accordingly to
\begin{equation}
\epsilon_i \rightarrow \epsilon_i + \delta \epsilon_i, \hspace{10pt} \delta \epsilon_i=  \mathcal{V}_{ii} \,,
\label{eq:pert_energy}
\end{equation}
leading to the perturbation in the argument of the interelectronic operator
\begin{equation}
I(\Delta_{va})\rightarrow I(\Delta_{va} + \delta \Delta_{va} ) \approx I(\Delta_{va}) + I^{\prime}(\Delta_{va}) (\delta \epsilon_v - \delta \epsilon_a)\,. 
\label{eq:pert_argument}
\end{equation}
The notation $\Delta_{va}$ stands for $\Delta_{va}= \epsilon_v - \epsilon_a$. The electron propagator (\ref{eq:e_propagator}) involved in the loops has to be perturbed as well. In its energy-position representation, one finds 
\begin{widetext}
\begin{align}
S_{\delta \mathcal{V}} (\epsilon_k \pm \omega; \textbf{x}, \textbf{y}) =  \delta_{\mathcal{V}} \left( \sum_i \frac{\ket{i} \bra{i}}{\epsilon_k \pm \omega- \epsilon_i u} \right) 
= \sum_{i,j} \frac{\ket{i} \mathcal{V}_{ij} \bra{j} }{ (\epsilon_k \pm \omega- \epsilon_i u) (\epsilon_k \pm \omega- \epsilon_j u) } - \sum_i \frac{ \ket{i} \mathcal{V}_{kk} \bra{i} }{ (\epsilon_k \pm \omega - \epsilon_i u) ^2}\,.
\label{eq:pert_propagator}
\end{align}
\end{widetext}
It is a slight generalization of the propagator found in Ref.~\cite{yerokhin.102.022815}. This expression is also valid when $\omega=0$, namely for the four-electron case. However, one should distinguish when to use each of the two terms. If both parentheses $(\epsilon_k \pm \omega - \epsilon_{i,j} u)|_{\omega=0}$ are non-zero, then the first piece of the perturbed propagator is to be used. If one of the parenthesis $(\epsilon_k \pm \omega - \epsilon_{i,j} u)|_{\omega=0}$ is zero, then the second piece of the perturbed propagator is to be used, with the appropriate index. In such a way, it reproduces the operator $\Xi$ of  Eq.~(46) in Ref.~\cite{Yerokhin_2021_104}. 

In the end, one trades the potential-link interaction matrix element for a one-photon-exchange one,
\begin{equation}
 \mathcal{V}_{ij}   \rightarrow  I_{i a j a}(0)   \,,
\label{eq:V_to_I}
\end{equation}
to match the three-photon-exchange formulas. Let us pause and comment on the last step presented above. A link between the potential-like interaction and the one-photon-exchange interaction is drawn. The logic goes as follows; consider for example, the Wichmann-Kroll potential \cite{Wichmann-Kroll:1956}. It accounts for an all-order treatment, in $\alpha Z$, of the interaction of the free-electron vacuum-polarization loop (first order in $\alpha$) with the Coulomb potential \cite{comment_Källen-Sabry}. 
\textcolor{red}{(}As a side comment, the Uehling potential \cite{Uehling:1935}, which takes into account the electric polarization of the vacuum (state), is the lowest order approximation (in $\alpha Z$) of the vacuum-polarization loop within the Coulomb field of the nucleus. One usually considers that the Wichmann-Kroll potential does not take the Uehling potential into account.\textcolor{red}{)} Albeit the Wichmann-Kroll potential is such a potential-like interaction, it deals with the full Green's function. Though, the potential-like interaction considered above incorporates only the core electrons. Therefore, going half a step back, one can imagine to have an all-order $\alpha Z$ Wichmann-Kroll-like potential mapped to the vacuum-polarization loop only for the core electrons. 
The inner structure of the "vacuum-polarization potential-like" interaction is then unravelled by a cut in the loop. Hence, one can think of the potential-like interaction as an effective one-electron operator, but implicitly accounting for a corrected wave function or representing a screened correction. 
\begin{figure}[h!]
    \centering
    \includegraphics[width=0.25\textwidth]{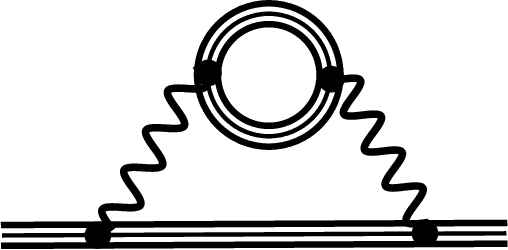}
    \caption{One-particle two-loop Feynman diagram corresponding to the perturbed S(VP)P subset in the redefined vacuum state formalism in an external potential $V$. The triple line indicates the electron propagator perturbed by the potential-like interaction $\mathcal{V}$. Notation is similar to Fig.~\ref{fig:H_subset}.} 
    \label{fig:perturbed_S(VP)E}
\end{figure}
%

Thus, one perturbs the S(VP)E equations (\ref{eq:S(VP)E,2e,irr})--(\ref{eq:S(VP)E,3e,red}) according to Eqs.~(\ref{eq:pert_state}--\ref{eq:pert_propagator}) and finds the desired formulas under the replacement (\ref{eq:V_to_I}). The Feynman diagram corresponding to the $\mathcal{V}$-perturbed S(VP)E subset is displayed in Fig.~\ref{fig:perturbed_S(VP)E}. The triple line notation represents the electron propagator perturbed by the potential-like interaction $\mathcal{V}$. It was shown that the two-electron subset and the three-electron subset are separately GI \cite{Soguel_2021pra}. The perturbation is initially a potential-like interaction $\mathcal{V}$, turned into a one-photon-exchange operator in the vanishing energy limit, which is gauge independent too. Therefore, it should be well-founded to expect that the results derived from this approach also fulfill the requirement of gauge invariance, namely on the three- and four-electron level respectively. The open question is whether the separation into S(VP)EVP and S[V(VP)P]E subsets holds also at the three- and four-electron level, as it is though to be the case. Hence, starting from the $\mathcal{V}$-perturbed S(VP)E GI expressions, one should be able to disentangle the various terms found via the redefinition of the vacuum state approach and assign them either to the three- or four-electron contribution. The results relying on this method are displayed in Appendix \ref{sec:third-order_3e} for the former and in Appendix \ref{sec:third-order_4e} for the latter.



\section{Regularization of infrared divergences}
\label{sec:IR_div}
Infrared divergences occur when the energy flowing through the loop ($\omega$) leads to a vanishing denominator of the electron propagator at $\omega\rightarrow0$. At the level of two-photon-exchange corrections, this behaviour was only met in ladder reducible expressions, where the removal of some terms in the summation of the crossed expressions was carried out to cancel IR divergent parts of the ladder reducible terms \cite{yerokhin:2001:032109}. In the case of the three-photon-exchange corrections, it can also arise from crossed reducible expressions. The analysis of IR divergences is conducted in the Feynman gauge, but the resulting pairing of the expressions is valid generally in virtue of gauge invariance. According to Shabaev \cite{shabaev:2002:119}, one introduces the following integral representation for the complex-exponential term of the photon propagator, including a photon mass term $\mu$,
\begin{equation}
    e^{i\sqrt{\omega^2 - \mu^2 + i \varepsilon} \lvert \textbf{x} - \textbf{y}\rvert} = \frac{-2}{\pi} \int_0^{\infty} dk \frac{k \sin(k \lvert \textbf{x} - \textbf{y}\rvert ) }{(\omega^2 - k^2 - \mu^2 + i\varepsilon)}\,.
\label{eq:massive_photon_propagtor}
\end{equation}
The photon mass plays the role of a energy cutoff (IR regulator). Notice that the condition on the branch of the square root is changed to Im$(\sqrt{\omega^2 - \mu^2 + i \varepsilon}) > 0$. In the rest of the section, the use of  $r_{12} \equiv \lvert \textbf{x} - \textbf{y}\rvert$ is preferred. Such type of integrals are met when facing IR divergences
\begin{equation}
\mathcal{I}_{n,m_{\pm},p} \equiv \frac{-i}{2\pi} \int d\omega \frac{I(\omega) I(\omega) I(0) }{ (-\omega + i \varepsilon)^{n} (\Delta \pm \omega + i\varepsilon )^m \widetilde{\Delta}^{p} } \,,
\label{eq:IR_div}    
\end{equation}
with $n=1,2,3$, $m=0,1$, $p=0,1$ and the constraint $n+m+p=3$. The $+$ sign stands for the IR divergent ladder reducible terms and the $-$ sign for the compensating crossed terms. Calculations are performed by the application of $\#$3.773(3) [or $\#$3.729(2)] in Gradshteyn and Ryzhik \cite{Gradshteyn_Ryzhik}. The first step is to show that for a first-order pole, $n=1$, the result is IR finite. There is no necessity to introduce a photon mass nor to use the integral representation of the complex exponential.  It is sufficient to Wick rotate the integration contour with the substitution $\omega = i\omega_E$.
Afterward, the principal value of the integral, denoted by $\mathcal{P}$, is considered. One shall start with the simplest case, 
%
\begin{align}
\mathcal{I}_{1,0,2} =&   \frac{ I(0) }{2\pi  \widetilde{\Delta}^2} \mathcal{P}\int_{0}^{i\infty} d\omega_E   \frac{ I(-i\omega_E) I(-i\omega_E) - I(i\omega_E) I(i\omega_E)}{ i\omega_E } \nonumber \\
&-  \frac{i}{ 2\widetilde{\Delta}^2 } I(0)I(0)I(0)    \,.  
\end{align}
%
The integral term reads explicitly 
\begin{equation}
\frac{\alpha^2}{r_{12}  r_{34} } \alpha_{1\mu} \alpha_2^{\mu} \alpha_{3\nu} \alpha_4^{\nu} \, \mathcal{P} \int_{0}^{i\infty} d\omega_E  \frac{  e^{-\omega_E R} - e^{\omega_E R}  }{ i\omega_E }  \,,  
\end{equation}
giving a finite contribution in the limit $\omega \rightarrow 0$. $R$ stands for $R= r_{12} + r_{34}$. One sees that the real part is IR finite, the imaginary one as well in this simple example. The second case, $\mathcal{I}_{1,2_-,0}$, requires one more step, a partial fraction decomposition, so that the Cauchy principal value can be applied. Recall that the principal value picks only the residues, hence no contribution from second or higher order poles.
\begin{widetext}
\begin{align}
\mathcal{I}_{1,2_-,0} =\, &  \frac{ I(0) }{2\pi \Delta^2} \mathcal{P} \int_0^{i\infty} d\omega_E  \left\{ \frac{ I(-i\omega_E) I(-i\omega_E) - I(i\omega_E) I(i\omega_E)}{ i\omega_E } - \frac{ I( -i \omega_E ) I( -i \omega_E  ) }{ \Delta + i\omega_E } -\frac{ I( i \omega_E ) I( i \omega_E  ) }{ \Delta - i\omega_E }  \right\} \nonumber \\
&-\frac{i}{ \Delta^2} I(0) \left[ I(0) I(0) - I(\Delta) I(\Delta) \right] \,.
\end{align} 
The first integrand was shown to IR finite just above. For the remaining two integrands, it is clear that $\Delta$ plays the role of an IR cutoff, preventing the expression to diverge in the limit $\omega \rightarrow0$. The case $\mathcal{I}_{1,1_-,1}$ is not met in the diagrams under consideration and is therefore not assessed.  Let us discuss the IR divergences arising from second-order poles. The first case \textcolor{red}{, $\mathcal{I}_{2,0,1}$,} considers the insertion of the interelectronic operator on the external leg of a one-loop diagram. One has, 
\begin{align}
\mathcal{I}_{2,0,1} = \frac{-\alpha^3 R}{\pi r_{12} r_{34} r_{56} \widetilde{\Delta}}\alpha_{1\mu} \alpha_2^{\mu} \alpha_{3\nu} \alpha_4^{\nu} \alpha_{5\rho} \alpha_6^{\rho} K_0(\mu R) 
\approx \frac{-\alpha^3 R}{\pi r_{12} r_{34} r_{56} \widetilde{\Delta}} \alpha_{1\mu} \alpha_2^{\mu} \alpha_{3\nu} \alpha_4^{\nu} \alpha_{5\rho} \alpha_6^{\rho} \left(\ln{\frac{\mu}{2}} + \gamma +\ln{R} \right)\,, 
\label{eq:div_ln(mu)}
\end{align}
in the limit $\mu \rightarrow 0$. $K_n(x)$ are imaginary Bessel functions of the second kind. The identical IR logarithmic divergent behaviour as in the two-photon-exchange ladder reducible case is recovered \cite{shabaev:2002:119}. The corresponding crossed diagram cancels the IR divergent term, leaving a well-behaved expression. The second case , $\mathcal{I}_{2,1_{+},0}$, is when the interelectronic operator is inserted within the electron propagator of the loop in the diagram. One faces 
\begin{align}
\mathcal{I}_{2,1_{+},0} =\, \frac{\alpha^3}{\pi r_{12} r_{34} r_{56} } \alpha_{1\mu} \alpha_2^{\mu} \alpha_{3\nu} \alpha_4^{\nu} \alpha_{5\rho} \alpha_6^{\rho}  \int_0^{\infty} dk  k \sin(kR) \left[ \frac{1}{(k^2 + \mu ^2 )^{3/2} (\Delta - \sqrt{k^2 + \mu ^2 } ) } - \frac{2}{(k^2 + \mu ^2 ) \Delta^2 }  \right] \,.
\end{align}
This IR divergence is compensated by a similar crossed graph, represented by the intergral $\mathcal{I}_{2,1_{-},0}$ where the interelectronic operator is also inserted within the electron propagator of the loop
\begin{align}
\mathcal{I}_{2,1_{-},0} = &\,\frac{-\alpha^3}{\pi r_{12} r_{34} r_{56}} \alpha_{1\mu} \alpha_2^{\mu} \alpha_{3\nu} \alpha_4^{\nu} \alpha_{5\rho} \alpha_6^{\rho} \int_0^{\infty} dk  k \sin(kR)  \nonumber \\
&\times \left[ \frac{-1}{(k^2  + \mu ^2 )^{3/2} (\Delta + \sqrt{k^2 + \mu ^2 } ) } + \frac{2}{(k^2 + \mu^2 - \Delta^2)\Delta^2  }- \frac{2}{(k^2 + \mu ^2 ) \Delta^2 }  \right]\,.
\end{align}
However, the cancellation is not straightforward at this step. A partial fraction decomposition allows to greatly simplify the previous expressions, when they are added together.
\begin{align}
\mathcal{I}_{2,1_{-},0} + \mathcal{I}_{2,1_{+},0} &= \frac{-2 \alpha^3}{\pi r_{12} r_{34} r_{56} \Delta^2 }  \alpha_{1\mu} \alpha_2^{\mu} \alpha_{3\nu} \alpha_4^{\nu} \alpha_{5\rho} \alpha_6^{\rho} \int_0^{\infty} dk \frac{k \sin(kR)} {k^2 + \mu^2 }\nonumber \\
& = \frac{-2 \alpha^3 }{\pi r_{12} r_{34} r_{56} \Delta^2 }  \alpha_{1\mu} \alpha_2^{\mu} \alpha_{3\nu} \alpha_4^{\nu} \alpha_{5\rho} \alpha_6^{\rho} \sqrt{\frac{\pi \mu R}{2}} K_{1/2} (\mu R)  \approx \frac{\alpha^3 }{r_{12} r_{34} r_{56} \Delta^2}  \alpha_{1\mu} \alpha_2^{\mu} \alpha_{3\nu} \alpha_4^{\nu} \alpha_{5\rho} \alpha_6^{\rho} \left( \mu R -1 \right) \,.
\end{align}
Thus, it turned out to be a spurious divergence; the IR divergence is ruled out and a finite part remains in the limit $\mu \rightarrow 0$. Last, but not least, the IR divergence arising from the third-order pole is treated. The corresponding integral to evaluate is
\begin{align}
\mathcal{I}_{3,0,0} &= \frac{-\alpha^3}{\pi r_{12} r_{34} r_{56} }  \alpha_{1\mu} \alpha_2^{\mu} \alpha_{3\nu} \alpha_4^{\nu} \alpha_{5\rho} \alpha_6^{\rho}\sqrt{\frac{\pi R }{2 \mu }} \frac{R}{2} K_{1/2} (\mu R) \approx \frac{-\alpha^3 R}{4  r_{12} r_{34} r_{56} } \alpha_{1\mu} \alpha_2^{\mu} \alpha_{3\nu} \alpha_4^{\nu} \alpha_{5\rho} \alpha_6^{\rho} \left( \frac{1}{\mu} - R \right)\,,
\label{eq:div_1/mu}
\end{align}
\end{widetext}
leading to a singular behaviour when the limit $\mu \rightarrow 0$ is taken. The behaviour of the IR divergence arising from the third-order pole is in agreement with the one presented in Ref.~\cite{yerokhin.102.022815}, where a similar analysis was performed for the self-energy screening effects in $g$-factor calculations. A different treatment for the regularization of the IR divergence for the third-order pole based on a symmetry argument is proposed in Appendix~\ref{appendix_ladred2}. Similarly to the IR divergence arising from the second-order pole, one looks for the compensating crossed diagram to ensure that the total expression is finite. 

The explicit cancellation of IR divergences, at the individual Feynman diagram level by the appropriate crossed term, is demonstrated in Table~\ref{tab:IR_div}. An exception is met for the ladder reducible 2 terms, which compensate themselves. A swapping of the indices in the expressions presented in Section~\ref{sec:third_order_rvs} might be sometimes necessary in order to make the compensation apparent. Moreover, it was shown that IR divergences are absorbed by expressions belonging to the same subset. 

\begin{table*}[]
\caption{IR divergences regularization at the individual Feynman diagram level. IR divergences are found in the reducible expressions, both for ladder- and crossed-loops. Crossed stands for the crossed compensating term. $\mathcal{I}_{n,m_\pm,p}$ describes the type of divergent integral encountered in the graph. Each term can be found in Chapter.~\ref{sec:third_order_rvs}, following the referenced equation. The labels are simplified in comparison to the ones presented there, only the difference among them is highlighted.}
\label{tab:IR_div}
\begin{center}
\tabcolsep10pt
\begin{tabular}{l | c | l | c }
\hline\hline 
$\Delta E^{(3\text{I})\text{3e}}_{v}$ $\in$ S[V(VP)P]E  &   &    &   \\
   & type of IR divergence & \multicolumn{1}{c|}{Crossed}  & IR compensation \\ \hline 
 
$H_1$: ladder red 1 (\ref{h1_lad_red1_10})   &  $\mathcal{I}_{2,1_+,0}$  &     $H_2 $: crossed (\ref{h2_cross})  &  $\mathcal{I}_{2,1_-,0}$  \\ [0.5ex] \hline 
 
$H_1$: ladder red 1(\ref{h1_lad_red1_11})  & $\mathcal{I}_{2,1_+,0}$   &    $H_2 $: crossed  (\ref{h2_cross}) & $\mathcal{I}_{2,1_-,0}$ \\ [0.5ex] \hline
 
 $H_1$: ladder red 2 (\ref{h1_lad_red2})      &$\mathcal{I}_{3,0,0}$     &$H_2$: ladder red 2  (\ref{h2_lad_red2} )   &  $\mathcal{I}_{3,0,0}$ \\ [0.5ex] \hline
 
 $H_1$: crossed red  (\ref{h1_cross_red1}) & $\mathcal{I}_{2,1_-,0}$ , $\mathcal{I}_{3,0,0}$   &    $H_2$: crossed    (\ref{h2_cross}) &  $\mathcal{I}_{2,1_-,0}$, $\mathcal{I}_{3,0,0}$\\ [0.5ex] \hline
 
 $H_2$: ladder red 1 (\ref{h2_lad_red1_24}) &   $\mathcal{I}_{2,0,1}$ &    $H_1 $: crossed irr  (\ref{h1_cross_irr_8}) & $\mathcal{I}_{2,0,1}$  \\ [0.5ex] \hline
 
 $H_2 $: ladder red 1 (\ref{h2_lad_red1_26}) &   $\mathcal{I}_{2,0,1}$&      $H_1 $: crossed irr   (\ref{h1_cross_irr_7})  & $\mathcal{I}_{2,0,1}$ \\ [0.5ex] \hline
 
 $H_2 $: ladder red 2 (\ref{h2_lad_red2}) &  $\mathcal{I}_{3,0,0}$   &    $H_1 $: ladder red 2 (\ref{h1_lad_red2})  &  $\mathcal{I}_{3,0,0}$ \\ [0.5ex] \hline
 
 $\Delta E^{(3\text{I})\text{3e}}_{v}$ $\in$ S(VP)EVP &   &    &   \\
 & type of IR divergence &  \multicolumn{1}{c|}{Crossed} & IR compensation \\ \hline 

 $F_1$: ladder red 1 (\ref{F1_lad_red12}) &  $\mathcal{I}_{2,0,1}$       & $F_1: $ crossed   (\ref{F1_cross}) & $\mathcal{I}_{2,0,1}$  \\[0.5ex]\hline

 $F_2$: ladder red 1(\ref{F2_lad_red11}) &  $\mathcal{I}_{2,1_+,0}$      & $F_2 $: crossed   (\ref{F2_cross}) & $\mathcal{I}_{2,1_-,0}$   \\[0.5ex]\hline
 
 $F_2$: ladder red 1 (\ref{F2_lad_red12}) &  $\mathcal{I}_{2,1_+,0}$       & $F_2 $ : crossed  (\ref{F2_cross})&  $\mathcal{I}_{2,1_-,0}$  \\[0.5ex]\hline
 
 $F_2$: ladder red 2 (\ref{F2_lad_red2}) &     $\mathcal{I}_{3,0,0}$   & $F_2 $: crossed   (\ref{F2_cross})&  $\mathcal{I}_{3,0,0}$ \\[0.5ex]\hline
 
 $F_3$: ladder red 1 (\ref{F3_lad_red12}) &  $\mathcal{I}_{2,0,1}$  &      $F_3 $: crossed irr   (\ref{F3_cross}) &  $\mathcal{I}_{2,0,1}$ \\[0.5ex] \hline
 
 $F_3$: ladder red 2 (\ref{F3_lad_red2}) &     $\mathcal{I}_{3,0,0}$        & $F_3 $: crossed red (\ref{F3_cross_red})&  $\mathcal{I}_{3,0,0}$ \\\hline
 
 $F_3$: crossed red (\ref{F3_cross_red}) &    $\mathcal{I}_{2,1_-,0}$       & $F_2 $ : crossed  (\ref{F2_cross}) &  $\mathcal{I}_{2,1_-,0}$\\
\hline\hline
\end{tabular}
\end{center}
\end{table*}

\section{Comparison}
\label{sec:comparison}

Two different approaches were utilized to infer a partial third-order interelectronic correction to the energy shift. A comparison between the results of each of these two approaches is undertaken in this section. 

The discussion begins with the four-electron contribution, which involves three types of terms: the irreducible [$\Delta E^{(3\text{I})\text{4e,irr}}_{v}$ (\ref{eq:4e_irr_F}, \ref{eq:4e_irr_H})], the reducible 1 [$\Delta E^{(3\text{I})\text{4e,red1}}_{v}$ (\ref{eq:4e_red1_F}, \ref{eq:4e_red1_H})] and the reducible 2 [$\Delta E^{(3\text{I})\text{4e,red2}}_{v}$ (\ref{eq:4e_red2_F}, \ref{eq:4e_red2_H})]. For every type previously stated, agreement between the perturbative treatment of the $\mathcal{V}$-perturbed S(VP)E three-electron subset and the effective one-particle approach is met. 
\newline \indent
For the three-electron contribution, an identical separation as above is conducted. The irreducible type is composed of three parts, two of which correspond to crossed diagrams, [$\Delta E^{(3\text{I})\text{3e,cross}}_v$ (\ref{eq:3e_irr_cross_H},\ref{eq:3e_irr_cross_F}), $\Delta E^{(3\text{I})\text{3e,cross,irr}}_v$ (\ref{eq:3e_irr_cross_irr_H}, \ref{eq:3e_irr_cross_irr_F})] and one corresponding to ladder diagrams [$\Delta E^{(3\text{I})\text{3e,lad,irr}}_v$ (\ref{eq:3e_irr_ladd_H}, \ref{eq:3e_irr_lad_F})]. The outcomes of the two methods are in full concordance for these terms. Regarding the reducible 1 type, the cross reducible [$\Delta E^{(3\text{I})\text{3e,cross,red}}_v$ (\ref{eq:3e_red_cross_H}, \ref{eq:3e_red_cross_F})] and the ladder reducible 1 IR free $\omega$  [$\Delta E^{(3\text{I})\text{lad,red1}}_{v,\,\text{IR free $\omega$}}$ (\ref{eq:3e_red1_lad_IRw_H}, \ref{eq:3e_red1_lad_IRw_F})] terms extracted from the two treatments are in good agreement. 
\newline
However, for the remaining reducible 1 terms [$\Delta E^{(3\text{I})\text{3e,lad,red1}}_{v,\,\text{IR div}}$ (\ref{eq:3e_red1_lad_IRdiv_H}, \ref{eq:3e_red1_lad_IRdiv_F}), $\Delta E^{(3\text{I})\text{3e,lad,red1}}_{v,\,\text{IR free}}$ (\ref{eq:3e_red1_lad_IRfree_H}, \ref{eq:3e_red1_lad_IRfree_F})] a discrepancy is encountered. Yerokhin, in Ref.~\cite{Yerokhin_2021_104}, already pointed out that the perturbation theory approach runs into troubles to deal with reducible terms (See remarks below Eqs.~(32, 35, 37) in Ref.~\cite{Yerokhin_2021_104}). He is invoking gauge invariance to fix the problem. If proceeding as explained at the beginning of the section dedicated to perturbation theory approach, an identical problem to the one highlighted by Yerokhin is met, namely that the poles differ by the sign of the $i\varepsilon$ prescription, 
\begin{equation}
\frac{1}{(\omega + i\varepsilon)(-\omega + i\varepsilon)} \hspace{10pt} \text{versus} \hspace{10pt} \frac{1}{(\omega + i\varepsilon)^2} \hspace{5pt} \text{or} \hspace{5pt} \frac{1}{(-\omega + i\varepsilon)^2} \,,    
\end{equation}
when facing ladder reducible 1 terms, $\Delta E^{(3\text{I})\text{3e,lad,red1}}_{v,\,\text{IR div}}$ and $\Delta E^{(3\text{I})\text{3e,lad,red1}}_{v,\,\text{IR free}}$. The difference in the topology of the poles arises from unaccounted restrictions in the summations. Surprisingly, and possibly related to the topology problem associated with the poles, the two approaches differ regarding the extra terms
\begin{align}
\label{eq:extra_term_H}
\Delta E^{(3\text{I})\text{3e,red1}}_{v, \, \text{S[V(VP)P]E}} = \sum_{a, b, b_1, v_1, i}^{i\neq v} \frac{I_{v b b_1 v_1}(\Delta_{vb}) I_{v_1 a i a}(0) I_{i b_1 b v}(\Delta_{vb}) }{ (\epsilon_v - \epsilon_i)^2}\,,
\end{align} 
and
\begin{align}
\Delta E^{(3\text{I})\text{3e,red1}}_{v, \, \text{S(VP)EVP}} = - \sum_{a, b, b_1, v_1, i}^{i\neq b} \frac{I_{v b b_1 v_1}(\Delta_{vb}) I_{b_1 a i a}(0) I_{i v_1 v b}(\Delta_{vb}) }{ (\epsilon_b - \epsilon_i)^2} \,.
\label{eq:extra_term_F}
\end{align}
These are not found via the perturbative analysis but are present in the redefined vacuum state approach. They are obtained as the interplay among terms generated from the ladder reducible 1 terms, upon the symmetrisation of the energy flow in the loop, and four-electron reducible 1 terms. The former originates from $H_1$ while the latter originates from $F_2$. They look like four-electron reducible 1 contribution, due to the absence of the  $\omega$ integration. According to the gauge invariance of the three-electron S(VP)E subset of the two-photon-exchange corrections, they should be incorporated in the three-electron contribution of the three-photon-exchange corrections. The structure of these terms suggests them to be included in the reducible 1 contribution. The last point to be made concerning these two terms is that they are obviously IR finite.   
\newline\indent
The reducible 2 type contains only ladder reducible 2 terms, the IR free one [$\Delta E^{(3\text{I})\text{3e,lad,red2}}_{v,\,\text{IR free }}$ (\ref{eq:3e_red2_lad_IRfree_H}, \ref{eq:3e_red2_lad_IRfree_F})] , and the IR divergent one [$\Delta E^{(3\text{I})\text{3e,lad,red2}}_{v,\,\text{IR div }}$ (\ref{eq:3e_red2_lad_IRdiv_H}, \ref{eq:3e_red2_lad_IRdiv_F})]. The expressions obtained from  the two different methods are identical.

Overall, if errors occurred in the different three-electron types of interelectronic interactions considered, one would expect to see repercussions in the four-electron contribution. In this case, the four-electron contribution would suffer from discrepancies between the two methods, a behaviour which is not encountered. Therefore, these two independent derivations and the comparison of the resulting expressions presented in the present work are fully consistent, except for two expressions mentioned above. In these cases, the discrepancy can be traced back to unaccounted restrictions in the summations, resulting in a different topology of the poles. Furthermore, the derivation based on this two-method scheme is a good sanity check of the obtained formulas. The perturbation theory approach also helped to sort out the three- and four-electron contributions, especially for the extra terms given in Eqs.~(\ref{eq:extra_term_H}, \ref{eq:extra_term_F}). In summary, the gathered evidences point towards the consistency of the derived three-electron expressions. Moreover, the excellent agreement met at the four-electron contribution level serves as a strong indication that the three-electron expressions should be reliable.  
\newline
Concerning the separation into the proposed GI subsets, S[V(VP)P]E and S(VP)EVP, the cancellation of IR divergences by elements from the same subset is very assuring. A numerical evaluation of the derived expressions is the sole way to either infirm the conjectured separability into the GI subsets put forward here, based on analytical considerations, or to confirm it and therewith turn it into a solid claim.

\section{Discussion and conclusion}
\label{sec:discussion}
The expressions presented in the work can readily be applied to any Li-like and/or B-like atoms or ions. They represent an important first step toward the evaluation of third-order interlectronic interactions. The three-photon-exchange formulas were explicitly derived for two proposed GI subsets arising from two classes of one-particle three-loop Feynman diagrams. The derivation relies on an effective one-particle approach, owing to the redefinition of the vacuum state, in the framework of the TTGF method. The resulting formulas contained infrared divergences. They were investigated and regularized by the introduction of a photon mass term. Two different types of divergences were encountered when the photon mass is sent to zero: a logarithmic one (\ref{eq:div_ln(mu)}) and a first-order singular one (\ref{eq:div_1/mu}). The divergent behaviours observed are in full accordance with previous studies \cite{shabaev:2002:119,yerokhin.102.022815}. In order to allow for a verification of the expressions derived in the framework of the TTGF, an independent derivation was conducted with the help of perturbation theory.  However, from the very beginning of the study, the issue of the topology of the pole was known for the second method; see the detailed explanation in the work of Yerokhin \textit{et al.} \cite{Yerokhin_2021_104}. Nevertheless, the idea of the present work was to see how far one can get with the cross-check, relying on the possibility of a perturbative treatment. This helped to resolve the different reducible terms (red1 and red2) and to sort out the distribution of different contributions (three-electron and four-electron) in each expression of the subsets (S[V(VP)P]E and S(VP)EVP). The agreement between the two approaches, at the four-electron level, is excellent. 
At the three-electron level, a reasonable agreement is met between the two results.   
Here, the discrepancy related to the different topology of the poles encountered in the (three-electron) ladder reducible 1 terms, for both IR divergent and IR finite ones, prevents to achieve a complete agreement. The difference in the topology of the poles could be attributed to unaccounted restrictions in summations. In fact such discrepancy in the topology of the poles was already encountered when comparing the results for the two-photon-exchange contributions between Refs.~\cite{sapirstein:2011:012504, sapirstein:2015:062508} and Ref.~\cite{Soguel_2021pra}. The comparison between the results for the two-photon-exchange two-electron contributions showed that the difference in the topology of the poles did not affect significantly the numerical values
; the difference amounts to ${\cal O} \left(7 \cdot 10^{-4}\right)$ atomic unit, or ${\cal O} \left( 1.9\cdot 10^{-2}\right)$ eV. An open question lies in the fact if the extra terms are related to this issue, and if the role they are playing potentially is to ensure gauge invariance. 

Based on the intuition acquired within the perturbative treatment, we do believe that the strong constraint of gauge invariance can be tracked to the third-order as well. Hence, it should be well-based to expect that the results derived from this approach also fulfill the requirement of gauge invariance.  Thus, in analogy to the perturbation theory approach, our belief is that the outcome of the TTGF method is also characterized by the important paradigm of gauge invariance.
Whether a further separation according to the subsets is possible cannot be resolved yet. Nevertheless, from the TTGF perspective, the explicit cancellation of IR divergences, within each subset, is very engaging. Last, but not least, a successful verification of the derived expressions was also carried out for the three-electron contribution with the $g$-factor formulas derived in Ref.~\cite{volotka:2014:253004}, under the replacement (\ref{eq:V_to_I}). Note that the extra terms are also present in those formulas; they manifest themselves when one proceeds towards numerical evaluations \cite{volotka:private}.

To conclude, the method based on a vacuum state redefinition in QED has shown, in this work, to be a well-suited tool to perform elaborate calculations. In contrast to other methods, it permits the identification of GI subsets and inherently, thus, validates the consistency of the obtained results. This asset can be very useful in future derivations of higher-order contributions since it provides a robust verification. Developing on this intrinsic characteristic, and to highlight the possibility to apply the formalism for advanced calculations, an investigation of third-order interelectronic corrections was carried out. Moreover, the identification of GI expressions within this approach paves the way for calculating higher-order corrections, which can be split into GI subsets and tackled one after the other. The presented redefined vacuum state approach can be further employed for atoms with a (more) sophisticated electronic structure, as it allows to focus only on the particles that differentiate between the configurations.

%

\section{Acknowledgments}
This work is supported by Bundesministerium für Bildung und Forschung (BMBF) through project 05P21SJFAA. R.N.S. is grateful to A. V. Volotka for helpful discussions, and to F. Karbstein for his help in improving the quality of the manuscript and its readability.

\appendix

\begin{widetext}
\section{Three-electron terms arising from the $H_2$ diagram}
\label{SVVPPE}

The expressions extracted from the $H_2$ Feynman diagram are found as follows,
\begin{equation} 
\Delta E^{(3\text{I}) \text{3e,cross}}_{v,\,H_2} = \frac{i}{2\pi} \int d\omega \sum_{i,j,k}  \frac{I_{v j i b}(\omega)  I_{k a j a}(0)  I_{b i k v}(\omega)  }{ (\epsilon_v -\omega -\epsilon_i u) (\epsilon_b -\omega -\epsilon_j u) (\epsilon_b -\omega -\epsilon_k u) }\,,
\label{h2_cross}
\end{equation}
\begin{equation} 
\Delta E^{(3\text{I})\text{3e,lad,irr}}_{v,\,H_2} = \frac{i}{2\pi} \int d\omega \sum_{i,j,k}^{k\neq b, \{i,j \} \neq \{b,v \} }  \frac{I_{v b i j }(\omega)  I_{k a b a}(0)  I_{j i k v}(\omega)  }{ (\epsilon_v -\omega -\epsilon_i u) (\epsilon_b +\omega -\epsilon_j u) (\epsilon_b -\epsilon_k u) }\,,
\end{equation}
\begin{equation} 
\Delta E^{(3\text{I}) \text{3e,lad,red1}}_{v,\,H_2} = -\frac{i}{2\pi} \int d\omega \sum_{i,j,k}^{k\neq b, \{i,j \} = \{b ,v \} }  \frac{I_{v b i j }(\omega)  I_{k a b a}(0)  I_{j i k v}(\omega)  }{ (\epsilon_v -\omega -\epsilon_i u)^2 (\epsilon_b -\epsilon_k u) }\,,
\label{h2_lad_red1_24}
\end{equation}
\begin{equation} 
\Delta E^{(3\text{I}) \text{3e,lad,irr}}_{v,\,H_2} = \frac{i}{2\pi} \int d\omega \sum_{i,j,k}^{k\neq b, \{i,j \} \neq \{b,v \} }  \frac{I_{v k i j }(\omega)  I_{ b a k a}(0)  I_{j i b v}(\omega)  }{ (\epsilon_v -\omega -\epsilon_i u) (\epsilon_b +\omega -\epsilon_j u) (\epsilon_b -\epsilon_k u) }\,,
\end{equation}
\begin{equation} 
\Delta E^{(3\text{I})\text{lad,red1}}_{v,\,H_2} = -\frac{i}{2\pi} \int d\omega \sum_{i,j,k}^{k\neq b, \{i,j \} = \{b,v \} }  \frac{I_{v k i j }(\omega)  I_{ b a k a}(0)  I_{j i b v}(\omega)   }{ (\epsilon_v -\omega -\epsilon_i u)^2 (\epsilon_b -\epsilon_k u) }\,.
\label{h2_lad_red1_26}
\end{equation}
The non-diagrammatic term for $H_2$ are found to be 
\begin{equation}
\Delta E^{(3\text{I})\text{3e,lad,red1}}_{v,\,H_2} = -\frac{i}{2\pi}\int d\omega  \sum_{i,j}^{\{i,j \} \neq \{b,v \} } \frac{I_{v b i j }(\omega)  I_{b_1 a b a}(0)  I_{j i b_1 v}(\omega)  }{ (\epsilon_v -\omega -\epsilon_i u) (\epsilon_b + \omega - \epsilon_j u)^2 }\,,
\label{h2_lad_red1_disc}
\end{equation}
\begin{equation}
\Delta E^{(3\text{I})\text{3e,lad,red2}}_{v,\,H_2} = -\frac{i}{2\pi} \int d\omega \sum_{i,j}^{\{i,j \} = \{b,v \} }  \frac{I_{v b i j }(\omega)  I_{b_1 a b a}(0)  I_{j i b_1 v}(\omega)  }{ (\epsilon_v -\omega -\epsilon_i u)^3 }\,.
\label{h2_lad_red2}
\end{equation}

\section{Three-electron terms arising from the $F_1$ and $F_3$ diagrams}
\label{SVPEVP} 

Within this subset, the Green's function associated to the remaining diagrams reads

\begin{align}
\Delta g^{(3) F_1}_{\alpha,\, vv}(E) =&\, \frac{1}{(E - \epsilon_v)^2} \left( \frac{i}{2\pi} \right)^3 \sum_{i, j, k, l, p} \int d\omega dk_1 dk_2 \frac{ I_{v i j i}(0) }{[k_1 - \epsilon_{i} + i\varepsilon (\epsilon_i - E_\alpha^F)] [E - \epsilon_{j} + i\varepsilon (\epsilon_j - E_\alpha^F)] } \nonumber \\
&\times  \frac{ I_{j p k l}(\omega)  I_{k l v p}(\omega)   }{  [E - \omega - \epsilon_{k} + i\varepsilon (\epsilon_k - E_\alpha^F)] [k_2 -  \epsilon_{l} + i\varepsilon (\epsilon_l - E_\alpha^F)] [k_2 -\omega - \epsilon_{p} + i\varepsilon (\epsilon_p - E_\alpha^F)]  }  \,, 
\end{align}
for $F_1$, and
\begin{align}
\Delta g^{(3) F_3}_{\alpha,\, vv}(E) =&\,  \frac{1}{(E - \epsilon_v)^2} \left( \frac{i}{2\pi} \right)^3 \sum_{i, j, k, l, p} \int d\omega dk_1 dk_2 \frac{ I_{v k i j}(\omega) }{[E - \omega - \epsilon_{i} + i\varepsilon (\epsilon_i - E_\alpha^F)] [k_1 - \epsilon_{j} + i\varepsilon (\epsilon_j - E_\alpha^F)] } \nonumber \\
&\times \frac{  I_{i j l k}(\omega)  I_{l p v p }(0)  }{   [k_1 - \omega - \epsilon_{k} + i\varepsilon (\epsilon_k - E_\alpha^F)] [E -  \epsilon_{l} + i\varepsilon (\epsilon_l - E_\alpha^F)] [k_2  - \epsilon_{p} + i\varepsilon (\epsilon_p - E_\alpha^F)]}    \,.
\end{align}
for $F_3$. The extraction procedure is carried out for the $F_1$ Feynman diagram, which contributes as follow. The terms associated to the crossed graphs are
\begin{equation}
\Delta E^{(3\text{I}) \text{3e,cross,irr}}_{v,\,F_1} = \frac{i}{2\pi} \int d\omega \sum_{i,j,k}^{i\neq v}   \frac{ I_{v a i a}(0)  I_{i k j b}(\omega)  I_{b j k v }(\omega)    }{  (\epsilon_v -  \epsilon_i ) (\epsilon_v - \omega - \epsilon_j u)  ( \epsilon_b -\omega - \epsilon_k u) }\,,
\label{F1_cross}
\end{equation}
\begin{equation} 
\Delta E^{(3\text{I}) \text{3e,cross,red1}}_{v,\,F_1} = -\frac{i}{2\pi} \int d\omega \sum_{i,j}   \frac{ I_{v a v_1 a}(0)  I_{v_1 k j b}(\omega)  I_{b j k v}(\omega)    }{   (\epsilon_v - \omega - \epsilon_j u)^2  ( \epsilon_b -\omega - \epsilon_k u) }\,.
\label{F1_cross_red}
\end{equation}
The expressions corresponding to the ladder-loop graph read
\begin{equation} 
\Delta E^{(3\text{I}) \text{3e,lad,irr}}_{v,\,F_1} = \frac{i}{2\pi} \int d\omega \sum_{i,j,k}^{i\neq v, \{j, k \} \neq \{v,b \} }   \frac{ I_{v a i a}(0)  I_{i b j k}(\omega)  I_{j k v b}(\omega)    }{ (\epsilon_v -  \epsilon_i u) (\epsilon_v - \omega - \epsilon_j u)  ( \epsilon_b +\omega - \epsilon_k u)  }\,,
\end{equation}
\begin{equation} 
\Delta E^{(3\text{I}) \text{3e,lad,red1}}_{v,\,F_1} = -\frac{i}{2\pi} \int d\omega \sum_{i,j,k}^{ \{j, k \} \neq \{v,b \} }   \frac{ I_{v a v_1 a}(0)  I_{v_1 b j k}(\omega)  I_{j k v b}(\omega)    }{ (\epsilon_v - \omega - \epsilon_j u)^2  ( \epsilon_b +\omega - \epsilon_k u)  }\,,
\label{F1_lad_red11}
\end{equation}
\begin{align} 
\Delta E^{(3\text{I}) \text{3e,lad,red1}}_{v,\,F_1} =& \,-\frac{i}{2\pi} \int d\omega \sum_{i,j,k}^{i\neq v, \{j, k \} = \{v,b \} }  \left\{ \frac{ I_{v a i a}(0)  I_{i b j k}(\omega)  I_{j k v b}(\omega)    }{ (\epsilon_v -  \epsilon_i u)^2 } \left[ \frac{1}{(\epsilon_v - \omega - \epsilon_j u)}  +   \frac{1}{(\epsilon_b +\omega - \epsilon_k u)} \right] \right.  \nonumber \\
&+ \left. \frac{ I_{v a i a}(0)  I_{i b j k}(\omega)  I_{j k v b}(\omega)  }{ (\epsilon_v -  \epsilon_i u)  (\epsilon_v - \omega - \epsilon_j u)^2    } \right\}\,,
\label{F1_lad_red12}
\end{align}
\begin{equation} 
\Delta E^{(3\text{I}) \text{3e,lad,red2}}_{v,\,F_1} = \frac{i}{2\pi} \int d\omega \sum_{j, k}^{ \{j, k \} = \{v,b \} }  \frac{ I_{v a v_1 a}(0)  I_{v_1 b j k}(\omega)  I_{j k v b}(\omega)    }{(\epsilon_v - \omega - \epsilon_j u)^3 }\,.
\label{F1_lad_red2}
\end{equation}
For the disconnected parts, the terms presented below cancel the reducible elements found in $F_1$, namely the term on the first line cancels (\ref{F1_cross_red}), the term in the second line cancels (\ref{F1_lad_red11}) and the term in the third line cancels (\ref{F1_lad_red2}). It leaves only the irreducible expressions and the ladder reducible 1 term (\ref{F1_lad_red12}).
\begin{align}
\Delta E^{(3\text{I}) \text{3e,disc}}_{v,\,F}=&\,  \frac{i}{2\pi} \int d\omega \left\{ \sum_{i,j}  \frac{ I_{v a v a}(0) I_{v_1 j i b}(\omega)  I_{i b v_1 j}(\omega) }{ (\epsilon_v - \omega-  \epsilon_i u)^2 (\epsilon_b - \omega - \epsilon_j u)  } + \sum_{i,j}^{ \{i,j\} \neq \{b,v\} }  \frac{ I_{v a v a}(0) I_{v_1 b i j}(\omega)  I_{i j v_1 b j}(\omega) }{ (\epsilon_v - \omega-  \epsilon_i u)^2 (\epsilon_b + \omega - \epsilon_j u)  } \right. \nonumber \\
&- \left.   \sum_{i,j}^{ \{i,j\} =\{b,v\} }  \frac{ I_{v a v a}(0) I_{v_1 b i j}(\omega)  I_{i j v_1 b j}(\omega) }{ (\epsilon_v - \omega-  \epsilon_i u)^3 } \right\} \,.
\end{align}
From the $F_3$ Feynman diagram arises the following terms. Similarly to the $F_1$ graph, the terms corresponding to the crossed graph are 
\begin{equation}
\Delta E^{(3\text{I}) \text{3e,cross,irr}}_{v,\,F_3} = \frac{i}{2\pi} \int d\omega \sum_{i,j,k}^{k \neq v} \frac{ I_{v j i b} (\omega) I_{i b k j} (\omega) I_{k a v a} (0) }{ (\epsilon_v - \omega - \epsilon_i u) (\epsilon_b - \omega - \epsilon_j u) (\epsilon_v  - \epsilon_k u) }\,,
 \label{F3_cross}
\end{equation}
\begin{equation} 
\Delta E^{(3\text{I}) \text{3e,cross,red}}_{v,\,F_3} = \frac{-i}{2\pi} \int d\omega \sum_{i,j} \frac{ I_{v j i b} (\omega) I_{i b v_1 j} (\omega) I_{v_1 a v a} (0) }{ (\epsilon_v - \omega - \epsilon_i u)^2 (\epsilon_b - \omega - \epsilon_j u)  }\,.
 \label{F3_cross_red}
\end{equation}
The ones associated to the ladder graph read
\begin{equation} 
\Delta E^{(3\text{I}) \text{3e,lad,irr}}_{v,\,F_3} = \frac{i}{2\pi} \int d\omega \sum_{i,j,k}^{\{i,j\} \neq \{b,v\}, k \neq v} \frac{ I_{v b i j} (\omega) I_{i j k b} (\omega) I_{k a v a} (0) }{ (\epsilon_v - \omega - \epsilon_i u) (\epsilon_b + \omega - \epsilon_j u) (\epsilon_v  - \epsilon_k u) }\,,
\end{equation}
\begin{equation} 
\Delta E^{(3\text{I}) \text{lad,red1}}_{v,\,F_3} = \frac{-i}{2\pi} \int d\omega \sum_{i,j}^{\{i,j\} \neq \{b,v\}} \frac{ I_{v b i j} (\omega) I_{i j v_1 b} (\omega) I_{v_1 a v a} (0) }{ (\epsilon_v - \omega - \epsilon_i u)^2 (\epsilon_b + \omega - \epsilon_j u)  }\,,
\label{F3_lad_red11}
\end{equation}
\begin{align}
\Delta E^{(3\text{I}) \text{3e,lad,red1}}_{v,\,F_3} =& \, \frac{-i}{2\pi} \int d\omega \sum_{i,j,k}^{\{i,j\} = \{b,v\}, k \neq v} \left\{ \frac{ I_{v b i j} (\omega) I_{i j k b} (\omega) I_{k a v a} (0) }{  (\epsilon_v  - \epsilon_k u)^2 } \left[   \frac{1}{ (\epsilon_v - \omega - \epsilon_i u) } + \frac{1}{ (\epsilon_b + \omega - \epsilon_j u) } \right] \nonumber  \right.\\
&+ \left. \frac{ I_{v b i j} (\omega) I_{i j k b} (\omega) I_{k a v a} (0) }{ (\epsilon_v - \omega - \epsilon_i u)^2 (\epsilon_v  - \epsilon_k u) } \right\}\,,
\label{F3_lad_red12}
\end{align}
\begin{equation} 
\Delta E^{(3\text{I}) \text{3e,lad,red2}}_{v,\,F_3} = \frac{i}{2\pi} \int d\omega \sum_{i,j}^{\{i,j\} = \{b,v\}} \frac{ I_{v b i j} (\omega) I_{i j v_1 b} (\omega) I_{v_1 a v a} (0) }{ (\epsilon_v - \omega - \epsilon_i u)^3  }\,.
\label{F3_lad_red2}
\end{equation}
\end{widetext}

\section{Third-order interelectronic corrections derived by a perturbative treatment: three-electron subset }
\label{sec:third-order_3e}
The idea would be to proceed as follows; one perturbs the S(VP)E two-electron expressions (\ref{eq:S(VP)E,2e,irr}, \ref{eq:S(VP)E,2e,red}) according to Eqs.~(\ref{eq:pert_state}--\ref{eq:pert_propagator}) and finds the desired formulas under the replacement (\ref{eq:V_to_I}). However, the issue is, as already pointed out by Yerokhin in Ref.~\cite{Yerokhin_2021_104}, that this approach suffers from troubles to deal with reducible terms \footnote{See \cite{comment_Yerokhin}.}. He is invoking gauge invariance to fix the problem. If proceeding as explained just above, an identical problem to the one highlighted by Yerokhin is met, namely that the poles differ by the sign of the $i\varepsilon$ prescription \footnote{See \cite{comment_Sapirstein}. },
\begin{equation}
\frac{1}{(\omega + i\varepsilon)(-\omega + i\varepsilon)} \hspace{10pt} \text{versus} \hspace{10pt} \frac{1}{(\omega + i\varepsilon)^2} \hspace{5pt} \text{or} \hspace{5pt} \frac{1}{(-\omega + i\varepsilon)^2} \,,    
\end{equation}
when facing ladder reducible 1 contributions, $\Delta E^{(3\text{I})\text{3e,lad,red1}}_{v,\,\text{IR div}}$ and $\Delta E^{(3\text{I})\text{3e,lad,red1}}_{v,\,\text{IR free}}$. The difference in the topology of the poles arises from  unaccounted restrictions in the summations. Nevertheless, for the sake of the verification, it is worth tackling this perturbative analysis and see how far one can get. However, due to this previous discrepancy, the different three-electron terms presented below are those derived within the redefinition of the vacuum state formalism. They are separated into irreducible (irr), reducible 1 (red1) and reducible 2 (red2) types, and moreover into S[V(VP)P]E and S(VP)EVP according to their originated diagrams.  
\begin{widetext}

\subsection{Irreducible terms }

The crossed irreducible expression, a new feature showing up at this order, is presented first. A separation according to the belonging to each subset is conducted,
%
\begin{equation}
\label{eq:3e_irr_cross_irr_H}
\Delta E^{(3\text{I})\text{3e,cross,irr}}_{v,\, \text{S[V(VP)P]E}} =  \frac{i}{2\pi} \int d \omega \sum_{a, b, i, j, k }^{k \neq b} \frac{ I_{v j i b}(\omega) I_{b a k a}(0) I_{i k v j}(\omega) + I_{v j i k}(\omega) I_{k a b a}(0) I_{i b v j}(\omega)  }{ (\epsilon_v - \omega - \epsilon_i u) (\epsilon_b - \omega - \epsilon_j u) \Delta_{bk} }\,,
\end{equation}
\begin{equation}
\Delta E^{(3\text{I})\text{3e,cross,irr}}_{v, \,\text{S(VP)EVP}} =  \frac{i}{2\pi} \int d \omega \sum_{a, b, i, j, k}^{i \neq v } \frac{ I_{v a i a}(0) I_{i k j b}(\omega) I_{b j k v}(\omega) + I_{v k j b}(\omega) I_{j b i k}(\omega) I_{a i a v}(0)  }{ \Delta_{vi}(\epsilon_v - \omega - \epsilon_j u) (\epsilon_b - \omega - \epsilon_k u)  }  \,.
\label{eq:3e_irr_cross_irr_F}
\end{equation}
%
Then, the crossed expression, also separated according to their origin diagrams, is displayed,
\begin{equation}
\label{eq:3e_irr_cross_H}
\Delta E^{(3\text{I})\text{3e,cross}}_{v,\, \text{S[V(VP)P]E}} =  \frac{i}{2\pi} \int d \omega \sum_{a, b, i, j, k}  \frac{  I_{v j i b}(\omega) I_{a k a j}(0) I_{i b v k}(\omega)   }{ (\epsilon_v - \omega - \epsilon_i u) (\epsilon_b - \omega - \epsilon_j u) (\epsilon_b - \omega - \epsilon_k u) } \,
\end{equation}
\begin{equation}
\Delta E^{(3\text{I})\text{3e,cross}}_{v,\, \text{S(VP)EVP}} =  \frac{i}{2\pi} \int d \omega \sum_{a, b, i, j, k}   \frac{  I_{v k i b}(\omega) I_{i a j a}(0) I_{b j k v}(\omega)  }{(\epsilon_v - \omega - \epsilon_i u) (\epsilon_v - \omega - \epsilon_j u) (\epsilon_b - \omega - \epsilon_k u)} \,.
\label{eq:3e_irr_cross_F}
\end{equation}
Finally, the ladder irreducible expression is found as 
\begin{align}
\label{eq:3e_irr_ladd_H}
\Delta E^{(3\text{I})\text{3e,lad,irr}}_{v, \, \text{S[V(VP)P]E}} =& \, \frac{i}{2\pi} \int d \omega \left[ \sum_{a, b, i, j, k}^{k \neq b, \{i,j \} \neq \{v,b \}} \frac{  I_{v b i j}(\omega) I_{a k a b}(0) I_{i j v k}(\omega) + I_{v k i j}(\omega) I_{a b a k}(0) I_{i j v b}(\omega)   }{ (\epsilon_v - \omega - \epsilon_i u) (\epsilon_b + \omega - \epsilon_j u) \Delta_{bk}} \right. \nonumber \\
&+ \left.\sum_{a, b, i,j,k}^{\{i,j \} \neq \{v,b \}, \{i,k \} \neq \{v,b \}}  \frac{ I_{v b i j}(\omega) I_{j a k a}(0) I_{i k v b}(\omega)   }{ (\epsilon_v - \omega - \epsilon_i u) (\epsilon_b + \omega - \epsilon_j u) (\epsilon_b + \omega - \epsilon_k u)} \right] \,, 
\end{align}
\begin{align}
\Delta E^{(3\text{I})\text{3e,lad,irr}}_{v, \, \text{S(VP)EVP}} =& \, \frac{i}{2\pi} \int d \omega \left[ \sum_{a, b, i, j, k}^{i \neq v, \{j,k \} \neq \{v,b \}} \frac{ I_{v a i a}(0) I_{i b j k}(\omega) I_{k j b v}(\omega) + I_{v b j k}(\omega) I_{j k i b}(\omega) I_{a i a v}(0)   }{  \Delta_{vi}(\epsilon_v - \omega - \epsilon_j u) (\epsilon_b + \omega - \epsilon_k u) } \right.\nonumber \\
&+ \left.  \sum_{a, b, i,j,k}^{\{i,k \} \neq \{v,b \}, \{j,k \} \neq \{v,b \}} \frac{ I_{v b i k}(\omega) I_{i a j a}(0) I_{k j b v}(\omega)  }{ (\epsilon_v - \omega - \epsilon_i u) (\epsilon_v - \omega - \epsilon_j u) (\epsilon_b + \omega - \epsilon_k u) } \right]\,.
\label{eq:3e_irr_lad_F}
\end{align}
\subsection{Reducible 1 terms }

Since a crossed irreducible expression exist, the associated crossed reducible terms are found and worked out. The result is separated as well according to the originated subset, and reads
\begin{equation}
\label{eq:3e_red_cross_H}
\Delta E^{(3\text{I})\text{3e,cross,red}}_{v,\, \text{S[V(VP)P]E}} = - \frac{i}{2\pi} \int d\omega \sum_{a, b, b_1, i,j} \frac{ I_{v j i b}(\omega) I_{b a b_1 a}(0) I_{i b_1 v j}(\omega)  }{ (\epsilon_v - \omega - \epsilon_i u)  (\epsilon_b - \omega - \epsilon_j u)^2}\,, 
\end{equation} 
\begin{equation}
\Delta E^{(3\text{I})\text{3e,cross,red}}_{v,\, \text{S(VP)EVP}} = - \frac{i}{2\pi} \int d\omega \sum_{a, b, v_1, i, j} \frac{ I_{v j i b}(\omega) I_{b i j v_1}(\omega) I_{a v_1 a v}(0)  }{ (\epsilon_v - \omega - \epsilon_i u)^2  (\epsilon_b - \omega - \epsilon_j u) }\,.
\label{eq:3e_red_cross_F}
\end{equation}
Each ladder reducible 1 term is separated into an IR free and an IR divergent part, and displayed according to its provenance. Beginning with the latter, one has
\begin{align}
\label{eq:3e_red1_lad_IRdiv_H}
\Delta E^{(3\text{I})\text{3e,lad,red1}}_{v,\,\text{S[V(VP)P]E} , \,\text{IR div}} = &\, -\frac{i}{2\pi} \int \frac{d\omega}{(-\omega +i\varepsilon)^2 } \sum_{a, b, b_1, v_1, i}^{i\neq b}  \left[  \frac{I_{v b v_1 b_1}(\omega) I_{a i a b}(0) I_{v_1 b_1 v i}(\omega) + I_{v i v_1 b_1 }(\omega) I_{a b a i}(0) I_{v_1 b_1 v b}(\omega) }{ \Delta_{bi}}  \nonumber \right.  \\
&+ \left. \frac{ I_{v b v_1 b_1}(\omega) I_{b_1 a i a}(0) I_{v_1 i v b}(\omega) + I_{v b v_1 i}(\omega) I_{i a b_1 a}(0) I_{v_1 b_1 v b}(\omega)  }{ (\Delta_{bi} + \omega + i\varepsilon)}  \right] \,, 
\end{align}
\begin{align}
\Delta E^{(3\text{I})\text{3e,lad,red1}}_{v,\,\text{S(VP)EVP} , \,\text{IR div}} =&\, -\frac{i}{2\pi} \int \frac{d\omega}{(-\omega +i\varepsilon)^2 }\sum_{a, b, b_1, v_1, i}^{i \neq v} \left[ \frac{  I_{v a i a}(0) I_{i b v_1 b_1}(\omega) I_{b_1 v_1 b v}(\omega) + I_{v b v_1 b_1}(\omega) I_{v_1 b_1 i b}(\omega) I_{a i a v}(0) }{\Delta_{vi} } \right. \nonumber \\
& \left. + \frac{ I_{v b i b_1}(\omega) I_{i a v_1 a}(0) I_{b_1 v_1 b v}(\omega) + I_{v b v_1 b_1 }(\omega) I_{v_1 a i a}(0) I_{b_1 i b v}(\omega)  }{ (\Delta_{vi} - \omega + i\varepsilon) }\right]\,,
\label{eq:3e_red1_lad_IRdiv_F}
\end{align}
for the IR divergent part. The IR finite part is further separated, because its first part is the counterpart of the previously introduced IR divergent terms,
\begin{align}
\label{eq:3e_red1_lad_IRfree_H}
\Delta E^{(3\text{I})\text{3e,lad,red1}}_{v,\, \text{S[V(VP)P]E},\,\text{IR free}} =&\, -\frac{i}{4\pi} \int d\omega \left[ \frac{1}{(\Delta_{vb} - \omega + i\varepsilon)^2} + \frac{1}{(\Delta_{vb} - \omega - i\varepsilon)^2}\right] \nonumber \\
&\times  \left[ \sum_{a, b, b_1, v_1, i}^{i \neq v }  \frac{ I_{v b b_1 v_1}(\omega) I_{v_1 a i a}(0) I_{b_1 i v b}(\omega) + I_{v b b_1 i}(\omega) I_{i a v_1 a}(0) I_{b_1 v_1 v b}(\omega) }{ (\Delta_{bi} + \omega + i\varepsilon)}  \right. \nonumber \\
&+ \left.  \sum_{a, b, b_1, v_1, i}^{i\neq b}   \frac{ I_{v b b_1 v_1}(\omega) I_{a i a b}(0) I_{b_1 v_1 v i}(\omega) + I_{v i b_1 v_1}(\omega) I_{a b a i}(0) I_{b_1 v_1 v b}(\omega) }{\Delta_{bi}}\right]  \,,
\end{align}
\begin{align}
\Delta E^{(3\text{I})\text{3e,lad,red1}}_{v,\, \text{S(VP)EVP},\,\text{IR free}} =&\, -\frac{i}{4\pi} \int d\omega \left[ \frac{1}{(\Delta_{vb} - \omega + i\varepsilon)^2} + \frac{1}{(\Delta_{vb} - \omega - i\varepsilon)^2}\right] \nonumber \\
&\times  \left[ \sum_{a, b, b_1, v_1, i}^{i\neq v}     \frac{ I_{v a i a}(0) I_{i b b_1 v_1}(\omega) I_{v_1 b_1 b v}(\omega) + I_{v b b_1 v_1}(\omega) I_{b_1 v_1 i b}(\omega) I_{a i a v}(0) }{\Delta_{vi}}   \right. \nonumber \\
&\left.  + \sum_{a, b, b_1, v_1, i}^{i\neq b}  \frac{ I_{v b b_1 v}(\omega) I_{b_1 a i a}(0) I_{v_1 i b v}(\omega) + I_{v b i v_1}(\omega) I_{i a b_1 a}(0) I_{v_1 b_1 b v}(\omega) }{ (\Delta_{vi} - \omega + i\varepsilon)}\right]  \,,
\label{eq:3e_red1_lad_IRfree_F}
\end{align}
while its second part is simply the terms excluded in sums of certain diagrams,  
\begin{equation}
\label{eq:3e_red1_lad_IRw_H}
\Delta E^{(3\text{I})\text{3e,lad,red1}}_{v,\,\text{S[V(VP)P]E},\,\text{IR free $\omega$}} = - \frac{i}{2\pi}\int d\omega \sum_{a, b, b_1, i, j}^{\{i,j\} \neq \{v,b\}}  \frac{ I_{v b i j}(\omega) I_{a b_1 a b}(0) I_{i j v b_1}(\omega)  }{ (\epsilon_v - \omega - \epsilon_i u) (\epsilon_b + \omega - \epsilon_j u)^2 }\,,
\end{equation}
\begin{equation}
\Delta E^{(3\text{I})\text{3e,lad,red1}}_{v,\,\text{S(VP)EVP},\,\text{IR free $\omega$}}=  - \frac{i}{2\pi}\int d\omega \sum_{a, b, v_1, i,j}^{\{i,j\} \neq \{v,b\}}  \frac{ I_{v b i j}(\omega) I_{i j v_1 b}(\omega) I_{a v_1 a v}(0)   }{(\epsilon_v - \omega - \epsilon_i u)^2  (\epsilon_b + \omega - \epsilon_j u)}\,.
\label{eq:3e_red1_lad_IRw_F}
\end{equation}

\subsection{Reducible 2 terms }
Facing now the reducible 2 contribution, the identical separation into IR divergent and IR free terms is conducted, on the top of the distinction among the two different subsets. The IR divergent terms are
\begin{align}
\label{eq:3e_red2_lad_IRdiv_H}
\Delta E^{(3\text{I})\text{3e,lad,red2}}_{v,\,\text{S[V(VP)P]E},\,\text{IR div }} =&\, \frac{i}{2\pi} \int \frac{d\omega}{(-\omega + i\varepsilon)^3} \sum_{a, b, b_1, b_2, v_1} \left[ I_{v b v_1 b_1}(\omega) I_{b_1 a b_2 a}(0) I_{v_1 b_2 v b}(\omega) - I_{v b v_1 b_2}(\omega) I_{a b_1 a b}(0) I_{v_1 b_2 v b_1 }(\omega)\right] \,, 
\end{align}
\begin{align}
\Delta E^{(3\text{I})\text{3e,lad,red2}}_{v,\,\text{S(VP)EVP},\,\text{IR div }} =&\, \frac{i}{2\pi} \int \frac{d\omega}{(-\omega + i\varepsilon)^3} \sum_{a, b, b_1, v_1, v_2}  \left[  I_{v b v_1 b_1}(\omega) I_{v_1 b_1 v_2 b}(\omega) I_{a v_2 a v}(0) - I_{v b v_1 b_1}(\omega) I_{v_1 a v_2 a}(0) I_{b_1 v_2  b v}(\omega)\right] \,,
\label{eq:3e_red2_lad_IRdiv_F}
\end{align}
and the IR free ones are
\begin{align}
\label{eq:3e_red2_lad_IRfree_H}
\Delta E^{(3\text{I})\text{3e,lad,red2}}_{v,\,\text{S[V(VP)P]E},\,\text{IR free }} =&\, \frac{i}{4\pi} \int d\omega \left[ \frac{1}{(\Delta_{vb} - \omega + i\varepsilon)^3} + \frac{1}{(\Delta_{vb} - \omega - i\varepsilon)^3 }\right]  \left[ \sum_{a, b, b_1, v_1, v_2} I_{v b b_1 v_1}(\omega) I_{v_1 a v_2 a}(0) I_{b_1 v_2 v b}(\omega) \right. \nonumber \\
& \left.- \sum_{a, b, b_1, , b_2, v_1}  I_{v b b_2 v_1}(\omega) I_{a b_1 a b}(0) I_{b_2 v_1 v b_1}(\omega) \right] \,, 
\end{align}
\begin{align}
\Delta E^{(3\text{I})\text{3e,lad,red2}}_{v,\,\text{S(VP)EVP},\,\text{IR free }} =&\, \frac{i}{4\pi} \int d\omega \left[ \frac{1}{(\Delta_{vb} - \omega + i\varepsilon)^3} + \frac{1}{(\Delta_{vb} - \omega - i\varepsilon)^3 }\right]  \left[ \sum_{a, b, b_1, v_1, v_2} I_{v b b_1 v_1}(\omega) I_{b_1 v_1 v_2 b}(\omega) I_{a v_2 a v}(0) \right. \nonumber \\
&\left. - \sum_{a, b, b_1, b_2, v_1} I_{v b b_1 v_1}(\omega) I_{b_2 v_1 v b}(\omega) I_{b_1 a b_2 a}(0) \right]  \,.
\label{eq:3e_red2_lad_IRfree_F}
\end{align}
\end{widetext}

\subsection{Extra terms }

The remaining terms
\begin{align}
\label{eq:app_extra_term_H}
\Delta E^{(3\text{I})\text{3e,red1}}_{v, \, \text{S[V(VP)P]E}} = \sum_{a, b, b_1, v_1, i}^{i\neq v} \frac{I_{v b b_1 v_1}(\Delta_{vb}) I_{v_1 a i a}(0) I_{i b_1 b v}(\Delta_{vb}) }{ (\epsilon_v - \epsilon_i)^2}\,,  
\end{align} 
and
\begin{align}
\Delta E^{(3\text{I})\text{3e,red1}}_{v, \, \text{S(VP)EVP}} = - \sum_{a, b, b_1, v_1, i}^{i\neq b} \frac{I_{v b b_1 v_1}(\Delta_{vb}) I_{b_1 a i a}(0) I_{i v_1 v b}(\Delta_{vb}) }{ (\epsilon_b - \epsilon_i)^2} \,,
\label{eq:app_extra_term_F}
\end{align}
are not found via the perturbative analysis but are present in the redefined vacuum state approach. They are obtained as the interplay among terms generated from the ladder reducible 1 terms, upon the symmetrization of the energy flow in the loop, and four-electron reducible 1 terms. The former originates from $H_1$ while the latter originates from $F_2$. They look like four-electron reducible 1 contribution, due to the absence of the  $\omega$ integration. According to the gauge invariance of the three-electron S(VP)E subset of the two-photon-exchange corrections, they should be incorporated in the three-electron contribution of the three-photon-exchange corrections. The structure of these terms suggests then to be included in the reducible 1 contribution.  

\section{Third-order interelectronic corrections derived by a perturbative treatment: four-electron subset }
\label{sec:third-order_4e}
One perturbs the S(VP)E three-electron expressions (\ref{eq:S(VP)E,3e,irr}, \ref{eq:S(VP)E,3e,red}) according to Eqs.~(\ref{eq:pert_state}--\ref{eq:pert_propagator}) and finds, under the replacement (\ref{eq:V_to_I}), three different types of four-electron contribution:  irreducible (irr), reducible 1 (red1) and reducible 2 (red2). They are displayed below, and furthermore separated into S(V(VP)P)E and S(VP)VP subsets. This separation relies on the redefined vacuum state analysis and its comparison with the perturbative one.

\begin{widetext}
\subsection{Irreducible terms }
The four-electron irreducible contribution is found to be
\begin{align}
\Delta E^{(3\text{I})\text{4e,irr}}_{v, \, \text{S(VP)EVP} } =& -\sum_{a, b, c, i, j}^{j\neq v, (i,c) \neq (v,b)} \left\{ \frac{    I_{a v a j}(0) \left[  I_{j b c i}(\Delta_{vc}) I_{c i v b}(\Delta_{vc}) + I_{j b i c}(\Delta_{cb}) I_{i c v b}(\Delta_{cb}) \right]}{ (\epsilon_v + \epsilon_b - \epsilon_c -  \epsilon_i) (\epsilon_v -  \epsilon_j)}  \right.\nonumber \\
&+  \left. \frac{\left[I_{c i j b}(\Delta_{vc}) I_{v b c i}(\Delta_{vc}) + I_{v b i c}(\Delta_{cb}) I_{i c j b}(\Delta_{cb}) \right] I_{j a v a}(0)   }{ (\epsilon_v + \epsilon_b - \epsilon_c -  \epsilon_i) (\epsilon_v -  \epsilon_j)}  \right\} \nonumber \\
&- \sum_{a, b, c, i, j}^{j\neq v} \frac{ I_{a v a j}(0) I_{j i c b}(\Delta_{vc}) I_{c b v i}(\Delta_{vc}) + I_{v i c b }(\Delta_{vc}) I_{c b j i }(\Delta_{vc}) I_{j a v a }(0) }{ (\epsilon_b + \epsilon_c - \epsilon_v -  \epsilon_i) (\epsilon_v -  \epsilon_j) } \nonumber \\
&- \sum_{a, b, c, i, j}^{j\neq c} \frac{ I_{v i j b}(\Delta_{vc}) I_{j a c a}(0) I_{c b v i}(\Delta_{vc}) + I_{v i c b }(\Delta_{vc}) I_{a  c a j }(0) I_{j b v i }(\Delta_{vc}) }{ (\epsilon_b + \epsilon_c - \epsilon_v -  \epsilon_i) (\epsilon_c -  \epsilon_j) } \nonumber \\
&-\sum_{a, b, c, i, j}^{j\neq c, (i,c) \neq (v,b)} \frac{ I_{a c a j}(0) I_{v b c i}(\Delta_{vc}) I_{j i v b}(\Delta_{vc})  + I_{v b j i}(\Delta_{vc}) I_{c i v b}(\Delta_{vc}) I_{j a c a}(0)  }{ (\epsilon_v + \epsilon_b - \epsilon_c -  \epsilon_i) (\epsilon_c -  \epsilon_j) } \nonumber \\
&- \sum_{i,j}  \frac{ I_{v i c b}(\Delta_{vc}) I_{i a j a}(0) I_{c b v j}(\Delta_{vc})    }{  (\epsilon_b + \epsilon_c - \epsilon_v -  \epsilon_i)  (\epsilon_b + \epsilon_c - \epsilon_v -  \epsilon_j)}   \,,
\label{eq:4e_irr_F}
\end{align}
for the S(VP)EVP subset and,
\begin{align}
\Delta E^{(3\text{I})\text{4e,irr}}_{v, \, \text{S[V(VP)P]E}} =&- \sum_{a, b, c, i, j}^{j\neq b, (i,c) \neq (v,b)} \left\{ \frac{ I_{a b a j}(0)\left[ I_{v j c i}(\Delta_{vc}) I_{c i v b}(\Delta_{vc}) + I_{v j i c}(\Delta_{cb}) I_{i c v b}(\Delta_{cb})   \right]}{ (\epsilon_v + \epsilon_b - \epsilon_c -  \epsilon_i) (\epsilon_b -  \epsilon_j)}  \right. \nonumber \\
&+ \left. \frac{\left[ I_{c i v j}(\Delta_{vc}) I_{v b c i}(\Delta_{vc}) + I_{v b i c}(\Delta_{cb}) I_{i c v j}(\Delta_{cb})   \right] I_{j a b a }(0) }{ (\epsilon_v + \epsilon_b - \epsilon_c -  \epsilon_i) (\epsilon_b -  \epsilon_j)}  \right\} \nonumber \\
&- \sum_{a, b, c, i, j}^{j\neq b} \frac{ I_{v i c j}(\Delta_{vc}) I_{j a b a}(0) I_{c b v i}(\Delta_{vc}) + I_{v i c b }(\Delta_{vc}) I_{a ba j }(0) I_{c j v i }(\Delta_{vc}) }{ (\epsilon_b + \epsilon_c - \epsilon_v -  \epsilon_i) (\epsilon_b -  \epsilon_j) } \nonumber \\
&- \sum_{a, b, c, i, j}^{ (i,c) \neq (v,b), (j,c) \neq (v,b) } \frac{ I_{v b c i}(\Delta_{vc}) I_{i a j a}(0) I_{c j v b}(\Delta_{vc}) + I_{v b i c}(\Delta_{cb}) I_{i a j a}(0) I_{j c v b}(\Delta_{cb}) }{ (\epsilon_v + \epsilon_b - \epsilon_c -  \epsilon_i) (\epsilon_v + \epsilon_b - \epsilon_c -  \epsilon_j) }  \nonumber \\ 
&-\sum_{a, b, c, i, j}^{j\neq c, (i,c) \neq (v,b)}  \frac{    I_{v b i j}(\Delta_{cb}) I_{i c v b}(\Delta_{cb})   I_{j a c a}(0) + + I_{a c a j}(0) I_{v b i c}(\Delta_{cb}) I_{i j v b}(\Delta_{cb})    }{ (\epsilon_v + \epsilon_b - \epsilon_c -  \epsilon_i) (\epsilon_c -  \epsilon_j) } \,, 
\label{eq:4e_irr_H}
\end{align}
for the S[V(VP)P]E one.

\subsection{Reducible 1 terms }
The reducible 1 contribution is also split into S(VP)EVP and S[V(VP)P]E subsets, respectively, and can be cast in the form 
\begin{align}
\Delta E^{(3\text{I})\text{4e,red1}}_{v, \, \text{S(VP)EVP}} =&-\sum_{a, b, b_1, v_1, i}^{i\neq b}  \frac{ I_{v b b_1 v_1}(\Delta_{vb})  I_{a b_1 a i}(0) I^{\prime}_{i v_1 v b}(\Delta_{vb})  + I_{a b a i }(0) I_{v i b_1 v_1}(\Delta_{vb}) I^{\prime}_{b_1 v_1 v b}(\Delta_{vb})   }{ (\epsilon_b - \epsilon_i) }  \nonumber \\
& - \sum_{a, b, b_1, v_1, i}^{i\neq v} \frac{ I_{v b b_1 i}(\Delta_{vb}) I_{i a v_1 a}(0)  I^{\prime}_{b_1 v_1 v b}(\Delta_{vb})  + I_{v b b_1 v_1}(\Delta_{vb}) I^{\prime}_{b_1 v_1 i b}(\Delta_{vb}) I_{i a v a}(0)     }{ (\epsilon_v - \epsilon_i)} \nonumber \\
&-\sum_{a, b, c, v_1, i}^{(i,c)\neq (v,b)}  \frac{ \left[ I_{v b c i}(\Delta_{vc}) I^{\prime}_{c i v b}(\Delta_{vc}) + I^{\prime}_{v b c i}(\Delta_{vc}) I_{c i v b}(\Delta_{vc}) \right]  I_{v_1 a v_1 a}(0)  }{(\epsilon_v + \epsilon_b - \epsilon_c -  \epsilon_i)}  \nonumber \\ 
&+\sum_{a, b, c, c_1, i}^{(i,c)\neq (v,b)}  \frac{ \left[ I_{v b c i}(\Delta_{vc}) I^{\prime}_{c i v b}(\Delta_{vc}) + I^{\prime}_{v b c i}(\Delta_{vc}) I_{c i v b}(\Delta_{vc}) \right]  I_{c_1 a c_1 a}(0) }{(\epsilon_v + \epsilon_b - \epsilon_c -  \epsilon_i)}  \nonumber \\ 
&- \sum_{a, b, c, v_1, i} \frac{ \left[I_{v i c b}(\Delta_{vc}) I^{\prime}_{c b v i}(\Delta_{vc}) + I^{\prime}_{v i c b}(\Delta_{vc})  I_{c b v i}(\Delta_{vc}) \right]  I_{v_1 a v_1 a}(0)   }{ (\epsilon_b + \epsilon_c - \epsilon_v -  \epsilon_i) }  \nonumber \\
&+ \sum_{a, b, c, c_1, i} \frac{ \left[I_{v i c b}(\Delta_{vc}) I^{\prime}_{c b v i}(\Delta_{vc}) + I^{\prime}_{v i c b}(\Delta_{vc})  I_{c b v i}(\Delta_{vc}) \right]  I_{c_1 a c_1 a}(0)    }{ (\epsilon_b + \epsilon_c - \epsilon_v -  \epsilon_i) }  \nonumber \\
&+\sum_{a, b, c, v_1, i}^{(i,c)\neq (v,b)} \frac{ I_{v_1 a v_1 a}(0) \left[ I_{v b c i}(\Delta_{vc}) I_{c i v b}(\Delta_{vc}) + I_{v b i c}(\Delta_{cb}) I_{i c v b}(\Delta_{cb}) \right]  }{ (\epsilon_v + \epsilon_b - \epsilon_c -  \epsilon_i)^2 } \nonumber \\
&- \sum_{a, b, c, v_1, i} \frac{  I_{v_1 a v_1 a}(0)  I_{v i c b}(\Delta_{vc}) I_{c b v i}(\Delta_{vc}) }{ (\epsilon_b + \epsilon_c - \epsilon_v -  \epsilon_i)^2 } \nonumber \\
&+ \sum_{a, b, c, c_1, i} \frac{  I_{c_1 a c_1 a}(0)   I_{v i c b}(\Delta_{vc}) I_{c b v i}(\Delta_{vc}) }{ (\epsilon_b + \epsilon_c - \epsilon_v -  \epsilon_i)^2 } \nonumber \\ 
&- \sum_{a, b, c, c_1, i}^{(i,c)\neq (v,b)} \frac{  I_{c_1 a c_1 a}(0) I_{v b c i}(\Delta_{vc}) I_{c i v b}(\Delta_{vc}) }{ (\epsilon_v + \epsilon_b - \epsilon_c -  \epsilon_i)^2 }  \,,
\label{eq:4e_red1_F}
\end{align}
and
\begin{align}
\Delta E^{(3\text{I})\text{4e,red1}}_{v, \, \text{S[V(VP)P]E}} =& - \sum_{a, b, b_1, v_1, i}^{i\neq v} \frac{ I_{a v a i}(0) I_{i b b_1 v_1}(\Delta_{vb})  I^{\prime}_{b_1 v_1 v b}(\Delta_{vb})  +    I_{v b b_1 v_1}(\Delta_{vb})  I_{a v_1 a i}(0) I^{\prime}_{b_1 i v b}(\Delta_{vb})  }{ (\epsilon_v - \epsilon_i)}  \nonumber \\
&-\sum_{a, b, b_1, v_1, i}^{i\neq b} \frac{   I_{v b b_1 v_1}(\Delta_{vb})  I^{\prime}_{b_1 v_1 v i}(\Delta_{vb}) I_{i a b a}(0)   +  I_{v b i v_1 }(\Delta_{vb}) I_{i a b_1 a}(0)   I^{\prime}_{b_1 v_1 v b}(\Delta_{vb}) }{ (\epsilon_b - \epsilon_i) } \nonumber \\
&-\sum_{a, b, c, c_1, i}^{(i,c)\neq (v,b)}    \frac{\left[ I_{v b i c}(\Delta_{cb}) I^{\prime}_{i c v b}(\Delta_{cb}) + I^{\prime}_{v b i c}(\Delta_{cb}) I_{i c v b}(\Delta_{cb})\right]  I_{c_1 a c_1}(0)  }{ (\epsilon_v + \epsilon_b - \epsilon_c -  \epsilon_i)}   \nonumber \\
&+\sum_{a, b, c, b_1, i}^{(i,c)\neq (v,b)}    \frac{\left[ I_{v b i c}(\Delta_{cb}) I^{\prime}_{i c v b}(\Delta_{cb}) + I^{\prime}_{v b i c}(\Delta_{cb}) I_{i c v b}(\Delta_{cb})\right]  I_{b_1 a b_1 a}(0)  }{ (\epsilon_v + \epsilon_b - \epsilon_c -  \epsilon_i)}   \nonumber \\
& + \sum_{a, b, c, b_1, i}^{(i,c)\neq (v,b)} \frac{ I_{b_1 a b_1 a}(0) \left[ I_{v b c i}(\Delta_{vc}) I_{c i v b}(\Delta_{vc}) + I_{v b i c}(\Delta_{cb}) I_{i c v b}(\Delta_{cb}) \right]  }{ (\epsilon_v + \epsilon_b - \epsilon_c -  \epsilon_i)^2 } \nonumber \\
&+ \sum_{a, b, c, b_1, i} \frac{  I_{b_1 a b_1 a}(0) I_{v i c b}(\Delta_{vc}) I_{c b v i}(\Delta_{vc})     }{ (\epsilon_b + \epsilon_c - \epsilon_v -  \epsilon_i)^2 } \nonumber \\
&- \sum_{a, b, c, c_1, i}^{(i,c)\neq (v,b)}  \frac{  I_{c_1 a c_1 a}(0) I_{v b i c}(\Delta_{cb}) I_{i c v b}(\Delta_{cb}) }{ (\epsilon_v + \epsilon_b - \epsilon_c -  \epsilon_i)^2 }  \,.
\label{eq:4e_red1_H}
\end{align}

\subsection{Reducible 2 terms }
Each of the inspected subset participates equally to the reducible 2 contribution. One finds 
\begin{align}
\Delta E^{(3\text{I})\text{4e,red2}}_{v, \, \text{S(VP)EVP}} =& -\frac{1}{2}  \sum_{a, b, b_1, v_1, v_2} \left[ I_{v b b_1 v_1} (\Delta_{vb}) I^{\prime \prime}_{b_1 v_1 v b} (\Delta_{vb}) + I^{\prime}_{v b b_1 v_1} (\Delta_{vb}) I^{\prime }_{b_1 v_1 v b} \right] I_{v_2 a v_2 a}(0) \nonumber \\
&+ \frac{1}{2}\sum_{a, b, b_1, b_2, v_1} \left[ I_{v b b_1 v_1} (\Delta_{vb}) I^{\prime \prime}_{b_1 v_1 v b} (\Delta_{vb}) + I^{\prime}_{v b b_1 v_1} (\Delta_{vb}) I^{\prime }_{b_1 v_1 v b} \right] I_{b_2 a b_2 a}(0) \,,
\label{eq:4e_red2_F}
\end{align}
originated from the $H$ diagrams, as well as
\begin{align}
\Delta E^{(3\text{I})\text{4e,red2}}_{v, \, \text{S[V(VP)P]E}} =& -\frac{1}{2}  \sum_{a, b, b_1, v_1, v_2} \left[ I_{v b b_1 v_1} (\Delta_{vb}) I^{\prime \prime}_{b_1 v_1 v b} (\Delta_{vb}) + I^{\prime}_{v b b_1 v_1} (\Delta_{vb}) I^{\prime }_{b_1 v_1 v b} \right] I_{v_2 a v_2 a}(0) \nonumber \\
&+ \frac{1}{2}\sum_{a, b, b_1, b_2, v_1} \left[ I_{v b b_1 v_1} (\Delta_{vb}) I^{\prime \prime}_{b_1 v_1 v b} (\Delta_{vb}) + I^{\prime}_{v b b_1 v_1} (\Delta_{vb}) I^{\prime }_{b_1 v_1 v b} \right] I_{b_2 a b_2 a}(0) \,,
\label{eq:4e_red2_H}
\end{align}
for the $F$ diagrams.

\end{widetext}

\section{Symmetry argument for vanishing ladder reducible 2 terms }
\label{appendix_ladred2}

The cancellation of the $1/\mu$ IR divergence in ladder reducible 2 terms was shown in Table~\ref{tab:IR_div}. This was achieved by searching for a compensating term within the subset of the investigated term. A symmetry argument might also rule out this issue at the individual Feynman diagram level, without the need of a term to absorb its divergence. The naive way to argue would be as follows.  Recall that in Feynman gauge, $I(\omega)$ is a symmetric operator. Therefore, when the IR divergence is met in the ladder reducible 2 term, one faces a symmetric numerator divided by an anti-symmetric denominator integrated over a symmetric interval. It vanishes due to parity consideration. However, the problem is that the $i\varepsilon$ prescription spoils the anti-symmetric behaviour of the denominator. Hence, a Wick rotation is applied and the expression looks as 
\begin{align}
\frac{i}{2\pi} &\int_{-\infty}^\infty d\omega \frac{I(\omega) I(\omega) I(0)}{ (-\omega + i\varepsilon)^3} = \nonumber \\
&  \frac{ -I(0)}{2\pi} \mathcal{P} \int_{0}^{i\infty} d\omega_E \frac{ I(-i\omega_E) I(-i\omega_E) - I(i\omega_E) I(i\omega_E)}{ (i\omega_E)^3} \propto  \nonumber \\
&  \frac{- i }{2\pi} I(0)  \mathcal{P} \int_{0}^{i\infty} d\omega_E \frac{ 2\sinh{ \omega_E R} }{ \omega_E^3 } \overset{\text{B.H}}{=} \nonumber \\
& \frac{-i R }{3\pi} I(0)  \mathcal{P} \int_{0}^{i\infty} d\omega_E \frac{ \cosh{\omega_E R }  }{ \omega_E^2 } \overset{\text{Taylor}}{\approx} \nonumber \\
&  \frac{-i R }{3\pi} I(0)  \mathcal{P} \int_{0}^{i\infty}  d\omega_E \left[ \frac{ 1}{\omega_E^2  } + \frac{1}{2!} R^2 \right]
\end{align}
Only the principal value is considered since the third order pole does not contribution to the pole term. One can take the exponential terms of the interelectronic operators out and rephrase them as a hyperbolic sine. Then, Bernoulli-L'Hospital rule is applied once and the hyperbolic cosine is Taylor expanded, as the interest lies in the low-energy limit. Here $R$ stands for $R=r_{12} + r_{34}$. An interesting feature is seen at this point, it leads to a pure imaginary end result. Furthermore, one retrieves the $1/\mu$ divergent behavior encountered previously. The imaginary contribution to the energy, the decay rate, accounts for the possible instability of the excited states. Since the interest lies in the energy level and not its lifetime, one can neglect it. For completeness, if the ground state is under consideration, it features obviously no instabilities. 

%

%

\bibliography{bibliography}
\end{document}